\documentclass[preprint,12pt]{elsarticle}

\usepackage{amssymb}
\usepackage{amsfonts}
\usepackage[toc,page]{appendix}
\usepackage[margin=2cm,bottom=4cm]{geometry}
\usepackage{subcaption}
\usepackage{pdfpages}
\usepackage{amsmath,bm}
\usepackage[noabbrev]{cleveref}
\usepackage{bbold}
\usepackage{caption}
\usepackage{amsthm}
\usepackage{url}
\usepackage{color,soul} 

\crefname{figure}{Fig.}{Figs.} \Crefname{figure}{Figure}{Figures} 
\crefname{equation}{Eq.}{Eqs.} \Crefname{equation}{Equation}{Equations}

\journal{Ultramicroscopy}

\begin{document}

\begin{frontmatter}



\title{Differential phase contrast from electrons that cause inner shell ionization}

\author[inst1]{Michael Deimetry}
\author[inst2]{Timothy C. Petersen}
\author[inst3]{Hamish G. Brown}
\author[inst2]{Matthew Weyland}
\author[inst1]{Scott D. Findlay\corref{cor1}}
\ead{scott.findlay@monash.edu}

\cortext[cor1]{Corresponding author}

\affiliation[inst1]{organization={School of Physics and Astronomy, Monash University},
            city={Clayton},
            state={Victoria},
            postcode={3800}, 
            country={Australia}}

\affiliation[inst2]{organization={Monash Centre for Electron Microscopy, Monash University},
            city={Clayton},
            state={Victoria},
            postcode={3800}, 
            country={Australia}}

\affiliation[inst3]{organization={Ian Holmes Imaging Center, University of Melbourne},
            city={Parkville},
            state={Victoria},
            postcode={3052}, 
            country={Australia}}

\begin{abstract}
Differential Phase Contrast (DPC) imaging, in which deviations in the bright field beam are in proportion to the electric field, has been extensively studied in the context of pure elastic scattering. Here we discuss differential phase contrast formed from core-loss scattered electrons, i.e. those that have caused inner shell ionization of atoms in the specimen, using a transition potential approach for which we study the number of final states needed for a converged calculation. In the phase object approximation, we show formally that differential phase contrast formed from core-loss scattered electrons is mainly a result of preservation of elastic contrast. Through simulation we demonstrate that whether the inelastic DPC images show element selective contrast depends on the spatial range of the ionization interaction, and specifically that when the energy loss is low the delocalisation can lead to contributions to the contrast from atoms other than that ionized. We further show that inelastic DPC images remain robustly interpretable to larger thicknesses than is the case for elastic DPC images, owing to the incoherence of the inelastic wavefields, though subtleties due to channelling remain. Lastly, we show that while a very high dose will be needed for sufficient counting statistics to discern differential phase contrast from core-loss scattered electrons, there is some enhancement of signal-to-noise ratio with thickness that makes inelastic DPC imaging more achievable for thicker samples.
\end{abstract}




\end{frontmatter}


\section{Introduction}

\begin{figure*}
\centering
\includegraphics[width=0.5\textwidth]{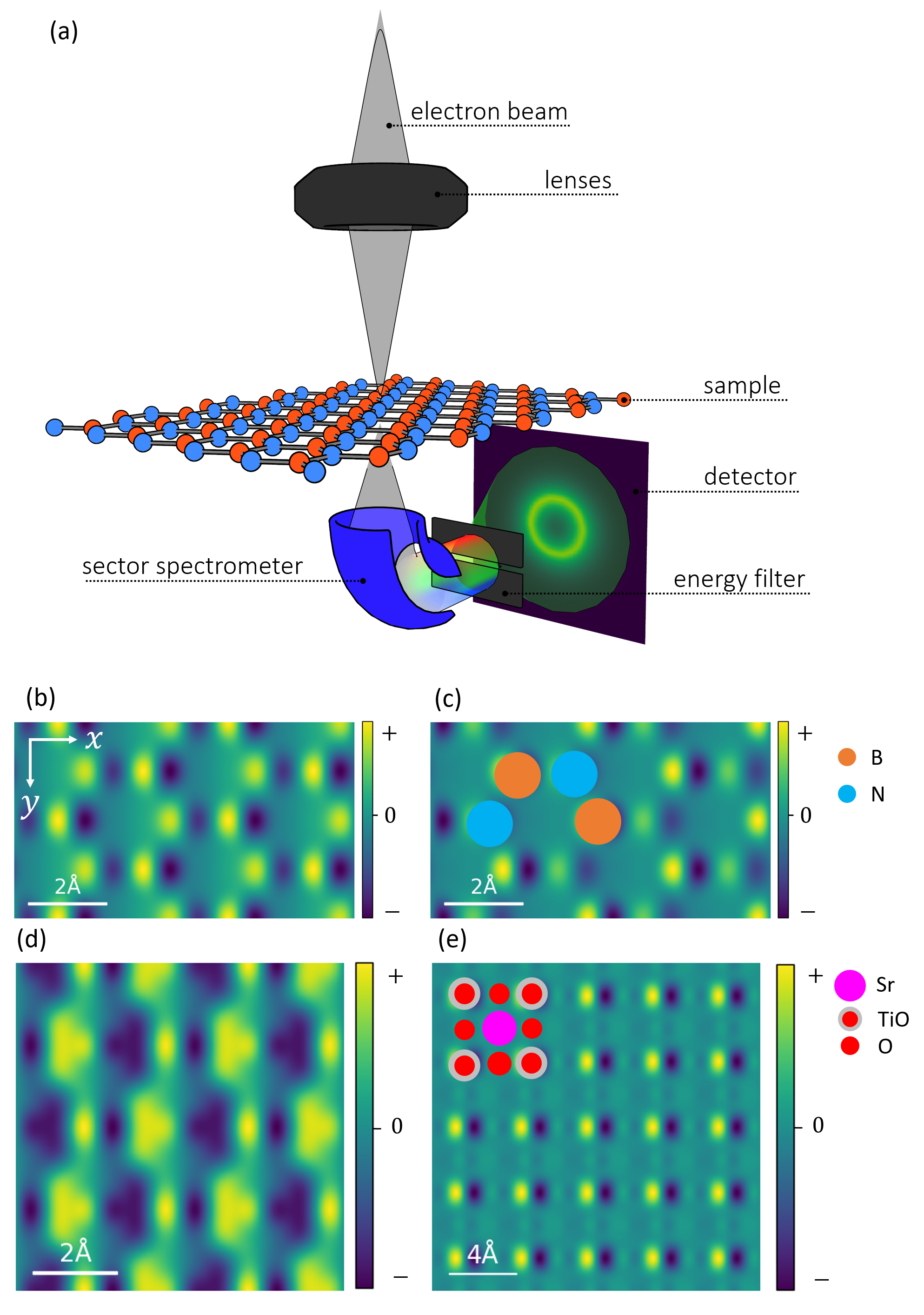}
    \hfill
    \caption{(a) Simplified schematic of the energy-filtered 4D STEM setup. An electron probe is raster scanned across a monolayer of hBN. The electrons pass through a spectrometer with a slit chosen to only let through electrons with kinetic energies in a desired energy range. (b) Elastic and (c) inelastic (with energy filter 50 eV above the boron K edge) simulated DPC images of the first moment along the $x$-direction (axes inset in (b)), for a detector spanning around $\pm$100 mrad in the $x$- and $y$-directions for monolayer hBN. The nitrogen columns are more pronounced in the elastic image, but still clearly visible in the inelastic image from the boron K shell. (d) Elastic and (e) inelastic (with energy filter 50 eV above the titanium L$_1$ edge) DPC images of the first moment along the $x$-direction for a detector with angular range extending out to 35 mrad for 200 {\AA} thick $\text{SrTiO}_3$. Whereas at this thickness it is difficult to interpret the structure clearly from the elastic image, the inelastic image is more robustly interpretable, with the Ti column locations showing clear differential phase contrast. These simulations assume 300 keV beam electrons and a 15.7 mrad probe-forming convergence semiangle. As the units for our EFDPC calculations are not intuitive, colour scales for EFDPC images simply show the zero value and distinguish positive from negative (see section \ref{sec:practical} for a discussion).
    }
    \label{fig:intro}
\end{figure*}

Fast readout electron pixel detectors are facilitating a range of momentum-resolved imaging modes now broadly referred to as four-dimensional scanning transmission electron microscopy\footnote{Not to be confused with STEM electron energy loss spectroscopy (EELS) tomography which is also referred to as 4D STEM EELS \cite{jarausch2009four}.} (4D STEM) \cite{ophus_four-dimensional_2019}, in which two-dimensional diffraction patterns are recorded at each position as a converged electron probe is raster scanned across the surface of a specimen. At nanometer resolution, this includes orientation mapping \cite{maclaren2020comparison,ophus2022automated}, strain measurement via nanobeam electron diffraction \cite{ozdol2015strain,shi2022uncovering}, and electromagnetic field mapping via differential phase contrast (DPC, also called centre-of-mass or first moment imaging)  \cite{tate2016high,krajnak2016pixelated,da2022influence}. At atomic resolution, this includes atomic-resolution DPC \cite{muller2014atomic,hachtel2018sub} and various forms of electron ptychography \cite{yang2016simultaneous,jiang2018electron,chen2021electron}. These techniques are predicated on the elastic scattering behaviour of electrons. Inelastic scattering is a confounding factor for these techniques, that can be mitigated if the 4D STEM data are zero-loss energy filtered \cite{bustillo20214d}. However, STEM with an energy filter before a fast-readout pixel detector, as depicted in \cref{fig:intro}(a), allows for additional imaging possibilities.

Simultaneous momentum and energy resolution has a long history in conventional TEM \cite{muto2017high}. In STEM, Midgley et al. \cite{midgley1995energy} showed a core-loss-filtered (i.e. from probe electrons that caused inner shell ionization events) convergent-beam electron diffraction pattern containing chemically-sensitive features in 1995. More recently, experimental realisation of simultaneous momentum and energy resolution in STEM by using an energy filter in front of a fast readout pixel detector has mainly focused on so-called $\omega$-$q$ imaging, resolving one direction of momentum transfer and energy loss to probe dispersion relations in plasmon and phonon scattering \cite{hage2017momentum,qi2021four}. Haas and Koch \cite{haas_momentum-_2022} used $\omega$-$q$ imaging with multiple slit orientations to synthesise core-loss-filtered DPC images of monolayer hexagonal boron nitride (hBN), showing atomic-resolution contrast qualitatively similar to that in zero-loss-filtered DPC images. While frameworks in which to simulate core-loss-filtered diffraction patterns are well established \cite{dwyer_multislice_2005,dwyer2008multiple,brown_linear-scaling_2019} – notably, Müller-Caspary et al. \cite{muller2017measurement} showed that simulated diffraction patterns in core-loss-filtered 4D STEM can contain rich structure that varies with probe position – there has been little exploration of how core-loss-filtered diffraction patterns might  be interpreted.

This paper explores the imaging formation dynamics of DPC imaging from core-loss-filtered 4D STEM,\footnote{All DPC images shown are calculated by evaluating the first moment assuming a pixel detector, though we expect results from a segmented detector to be similar, as is the case for elastic DPC imaging \cite{close_towards_2015,seki2017quantitative}.} describing core-loss scattering via the transition potential formulation \cite{dwyer_multislice_2005,brown_linear-scaling_2019,dalfonso_three-dimensional_2008,coene_inelastic_1990}. Our exploration is centred around two key results. The first is the experimental finding of Haas and Koch \cite{haas_momentum-_2022} that the zero-loss-filtered and core-loss-filtered DPC images have qualitatively similar appearance, as reproduced in the simulated DPC images for hBN in \cref{fig:intro}(b) and (c) respectively. In particular, the nitrogen atoms are clearly visible in the core-loss-filtered DPC image for the boron K edge, and so, in this case, core-loss-filtered DPC imaging is not providing element-specific contrast. The second, demonstrated in the simulated DPC images for 200 {\AA} thick SrTiO$_3$ in \cref{fig:intro}(d) and (e), is that the image contrast in core-loss-filtered DPC imaging (\cref{fig:intro}(e)) is more robust with respect to thickness (and defocus) than that in zero-loss-filtered DPC imaging (\cref{fig:intro}(d)). While \cref{fig:intro}(b)-(e) assume infinite dose to emphasise the similarity in appearance between elastic and core-loss-filtered DPC images, in practice core-loss-filtered DPC images will typically have contrast several orders of magnitude smaller than that for elastic DPC images. We will use infinite dose calculations to explore contrast formation mechanisms and imaging dynamics, but return to practical considerations about dose later in the paper.

A brief note on terminology is warranted. While there is a close analogy between energy-filtered transmission electron microscopy (EFTEM) and energy-filtered 4D STEM, as an acronym `EFSTEM' is already used in different contexts \cite{wang2011spatial}. For specificity, we will therefore refer to energy-filtered DPC, abbreviated to EFDPC, which unless otherwise specified should in this work be understood to mean core-loss filtering.

\section{Transition potential formulation of core-loss scattering}

\begin{figure}
    \centering
    \includegraphics[width=.4\textwidth]{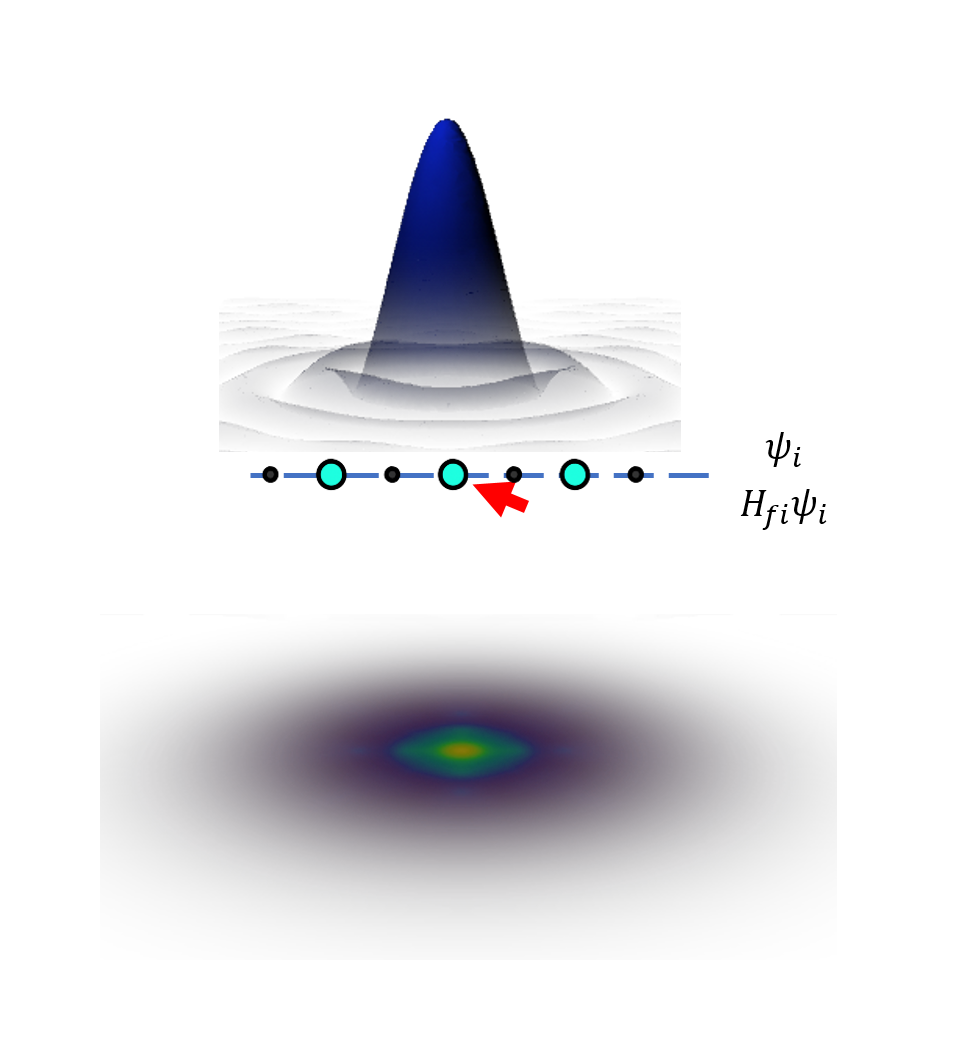}
    \caption{Schematic of the inelastic scattering process in a monolayer of atoms. The incident wavefield $\psi_i$ produces an inelastic wave governed by the transition potential $H_{fi}$. This procedure is repeated for each final state of each atom of the ionized species, and the resultant diffraction patterns are summed incoherently.}
    \label{fig:calc_schem_monolayer}
\end{figure}

To simulate core-loss-filtered 4D STEM data, we follow the approach of Coene and Van Dyck \cite{coene_inelastic_1990}, further developed for core-loss transitions by Dwyer \cite{dwyer_multislice_2005}. The basic concept is sketched in \cref{fig:calc_schem_monolayer} for a single layer of atoms. The inelastically-scattered wavefield $\psi_f$ that results when an incident elastic wave $\psi_i$ excites an atom from its initial state $i$ to some final state $f$ is given by \cite{dwyer_multislice_2005,dalfonso_three-dimensional_2008,coene_inelastic_1990}
 \begin{align}
     \psi_f(\textbf r_\perp) = -i \sigma H_{fi}(\textbf r_\perp) \psi_i(\textbf r_\perp) \;,
     \label{eq:Y_eqns_soln}
 \end{align}
where $\textbf{r}_\perp$ are the real-space coordinates transverse to the optic axis, $\sigma = \gamma m_0 \lambda / 2\pi \hbar^2$ (in which $\gamma$ is the Lorentz factor, $m_0$ is the electron rest mass, $\lambda$ is the relativistically-corrected de-Broglie wavelength and $\hbar$ is the reduced Planck's constant), and $H_{fi}(\textbf r_\perp)$ is a (projected) transition potential\footnote{In the literature $H_{fi}(\textbf r_\perp)$ is also referred to as a projected matrix element, M$\o$ller potential or simply a matrix element. The choice of units also varies, with other choices leading to a prefactor that differs accordingly from the $\sigma$ of \cref{eq:Y_eqns_soln}.}. Since for the single plane of atoms depicted in \cref{fig:calc_schem_monolayer} the wavefield $\psi_f(\textbf r_\perp)$ is at the exit surface of the sample, its contribution to the diffraction pattern is given by the intensity of its Fourier transform. The principles of quantum mechanics  oblige us to (incoherently) sum all contributions from sample final states that are not directly observed but which are consistent with what is measured. Thus the core-loss-filtered diffraction intensity may be written
 \begin{align}
     {\mathcal I}({\bf k}_\perp) = \sum_f \left| {\mathcal F}\{\psi_f\}({\bf k}_\perp) \right|^2 \;,
     \label{eq:inel_diff_patt}
 \end{align}
where the sum over $f$ includes all inner shell ionization final states consistent with the selected energy loss (in practice, rather an energy loss range), $ {\mathcal F}$ denotes Fourier transform, and ${\bf k}_\perp$ denotes the diffraction plane coordinate. 

Following Dwyer \cite{dwyer_multislice_2005} in assuming a central field, one-electron wavefunction model for the atomic electron being ejected through the ionization event, we label the initial states via the quantum numbers $n$ (principal quantum number), $\ell$ (orbital quantum number) and $m$ (magnetic quantum number), and the final states via the quantum numbers $\varepsilon$ (being the kinetic energy of the ejected electron, equal to the energy loss above the atom's ionization threshold), $\ell'$ and $m'$ (the orbital and magnetic quantum numbers, respectively). Throughout this paper, we adopt the convention of denoting continuum orbital and magnetic quantum numbers using primed variables. The three-dimensional transition potentials in reciprocal space are given by  \cite{dwyer_multislice_2005,dalfonso_three-dimensional_2008}
\begin{equation}
    H_{n\ell m \rightarrow \varepsilon\ell' m'}(\textbf k) =\frac{q_e^2}{4\pi^2 \varepsilon_0 k^2} \int d\textbf r \, a_{n\ell m}(\textbf r) \, a^*_{\varepsilon \ell' m'}(\textbf r) \, e^{-i 2\pi \textbf k \cdot \textbf r} \;,
    \label{eq:Hn0_k_space}
\end{equation}
where $q_e$ is the elementary charge and $\varepsilon_0$ is the permittivity of free space. \Cref{eq:Hn0_k_space} corresponds to the coupling of the bound state $ a_{n\ell m}(\textbf r)$ and the continuum state $a_{\varepsilon \ell' m'}(\textbf r)$ of the atomic electron ejected during the ionization event.\footnote{We will not consider bound-bound transitions, though they can be described by a similar formalism \cite{xin2014there}.} The projected transition potential appearing in \cref{eq:Y_eqns_soln} is then obtained by \cite{dwyer_multislice_2005,dalfonso_three-dimensional_2008}
 \begin{align}
      H_{fi}(\textbf r_\perp) &= \int_{-\infty}^{\infty} dz'\,  e^{-i 2 \pi (k_f - k_i) z'} H_{fi}(\textbf r_\perp, z') \\
      &= \int d\textbf k_\perp  H_{fi}(\textbf k_\perp, k_f - k_i) \, e^{i2\pi \textbf k_\perp \cdot \textbf r_\perp} \;,
 \end{align}
where the second equality follows by the Fourier projection theorem, and $k_f-k_i$ is related to the energy loss of the probe electron. We will refer to $H_{fi}$ simply as a transition potential, with the argument implicitly denoting whether the potential is projected or otherwise.

Writing the bound and continuum states explicitly as a product of angular and radial terms\footnote{Following Ref. \cite{oxley2000atomic}, we use relativistic Hartree-Fock wave functions for the bound states and Hartree-Slater wave functions for the continuum states.}, $a_{n\ell m}(\textbf r)=Y^{m}_{\ell}(\hat{\textbf{r}}) P_{n \ell}(r)/r$ and $a_{\varepsilon \ell' m'}(\textbf r)=Y^{m'}_{\ell'}(\hat{\textbf{r}}) P_{\varepsilon \ell'}(r)/r$, \cref{eq:Hn0_k_space} can be simplified to
\begin{align}
      H_{n\ell m \rightarrow \varepsilon\ell' m'}(\textbf k) = \frac{q_e^2}{4\pi^2 \varepsilon_0 k^2} \sum_{\ell''=|\ell'-\ell|}^{\ell'+\ell} (-i)^{\ell''} Y^{m-m'}_{\ell''}(\hat{\textbf{k}}) \langle l' m' | \overline{l''(m-m')}|lm\rangle R_{l',l'',l}(k) \;,
      \label{eq:Hfi_full_exp}
\end{align}
where the Gaunt coefficient
\begin{align}
 \langle l' m' | \overline{l''m''}|lm\rangle &= \int d \hat{\textbf{q}} \, Y^{m'*}_{\ell'}(\hat{\textbf{q}}) Y^{m''*}_{\ell''}(\hat{\textbf{q}}) Y^{m}_{\ell}(\hat{\textbf{q}}) \nonumber \\ &= (-1)^{m'+m''} \sqrt{\frac{(2\ell'+1)(2\ell''+1)(2\ell+1)}{4\pi}}
    \begin{pmatrix}
            \ell' & \ell'' & \ell \\
            0 & 0 & 0
            \end{pmatrix}
            \begin{pmatrix}
            \ell' & \ell'' & \ell \\
            -m' & -m'' & m
            \end{pmatrix}
            \label{eq:gaunt}
\end{align}
(in which the array-like objects are Wigner 3-j symbols), and 
\begin{align}
  R_{l',l'',l}(k) = \int_0^\infty dr P_{\varepsilon \ell'}(r) j_{\ell''}(k r) P_{nl}(r) = \sqrt{\frac{\pi}{2k}} \mathbb{HT}_{\ell''+1/2}\left\{ \frac{P_{\varepsilon \ell'}(r) P_{n\ell}(r)}{r^{3/2}} \right\}(k)
     \label{eq:overlap_int_shorthand}
\end{align}
(in which $j_{\ell''}$ are spherical Bessel functions of order $\ell''$ of the first kind) can be computed efficiently using the fast Hankel transform\footnote{The Hankel transform is defined as \begin{align*}
    \mathbb{HT}_{\mu}\{f\}(k) = \int_0^\infty dr \, f(r) J_{\mu}(kr) r \;,
\end{align*}
where $J_\mu$ is a $\mu^{\text{th}}$ order Bessel function of the first kind.}. Calculations in this paper are based on these ideas as implemented in py\_multislice \cite{pyms}.

\begin{figure}
    \centering
    \includegraphics[width=.45\textwidth]{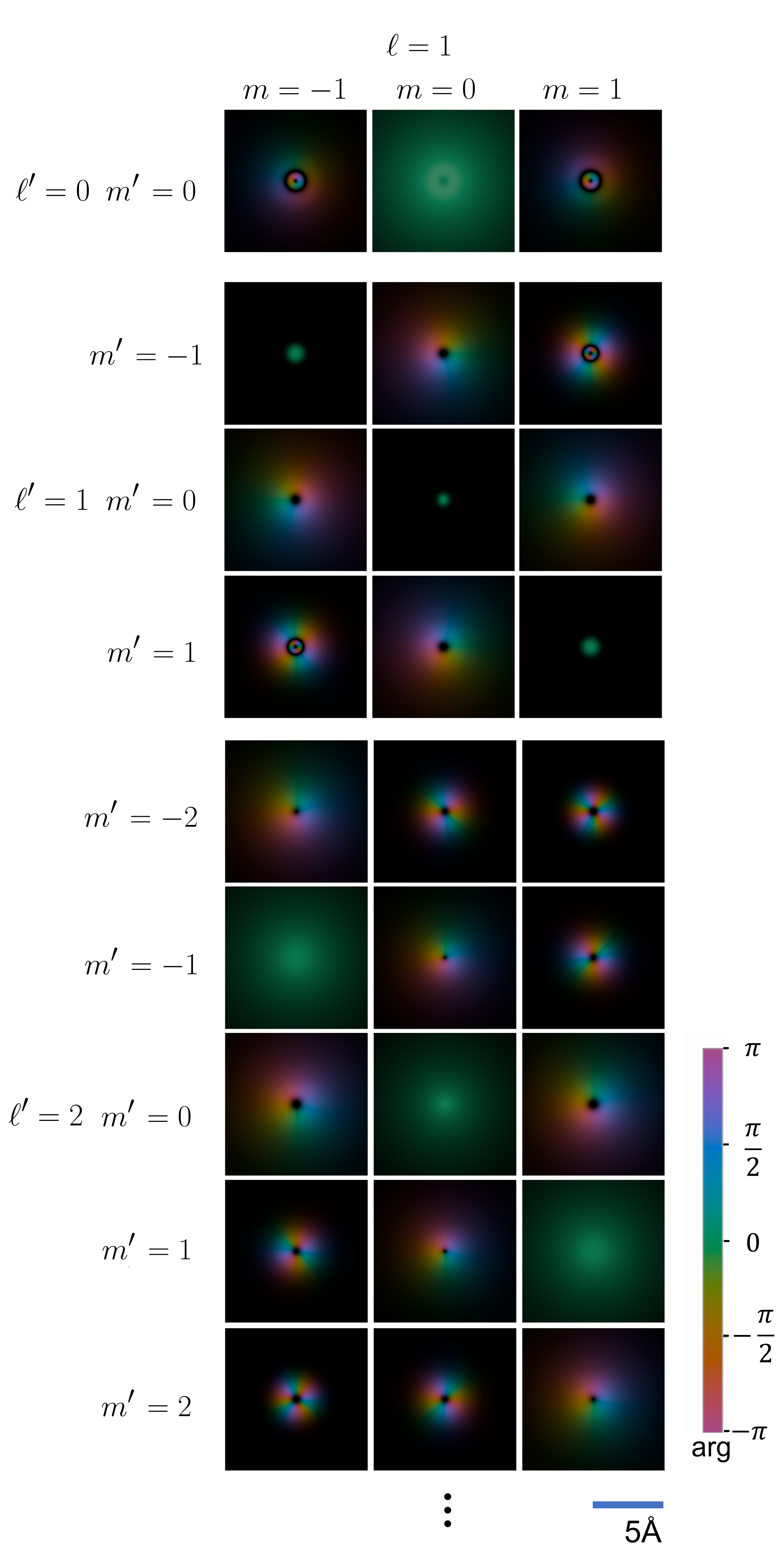}
    \caption{L$_{2,3}$ shell titanium transition potentials, $H_{n\ell m \rightarrow \epsilon \ell' m'}$, for beam energy 80keV and $\varepsilon=10$eV. The magnitude, depicted as brightness, shows that beam electrons near the atom are most likely to cause ionization events. The phase, depicted modulo $2\pi$ by colour, has a vortex structure with the winding number given by $m-m'$.}
    \label{fig:Hn0_plots}
\end{figure}

\Cref{fig:Hn0_plots} shows various titanium L$_{2,3}$ shell transition potentials. The greatest amplitude is near the atom, implying beam electrons closest to the titanium atom are more likely to cause ionization events. The vortex phase structure occurs because, despite the sum over $\ell''$ in \cref{eq:Hfi_full_exp}, all the azimuthal dependence is contained in a phase factor of the form $e^{i(m-m')\phi_k}$. The so-called winding number of the vortex is given by $m-m'$, where positive values imply increasing phase in the clockwise direction. The transitions are related by the magnetic quantum numbers through the relationship 
\begin{align}
H^*_{\ell(-m) \rightarrow \ell' (-m') }({\bf r_\perp}) = (-1)^{\ell'+\ell} H_{ \ell  m\rightarrow \ell' m' }({\bf r_\perp})
 \label{eq:Hfi_symmetry}
\end{align}
which is seen in \cref{fig:Hn0_plots}, where the tiles ($m$,$m'$) and ($-m$,$-m'$), located in the diametrically opposed position about the ($m=0$,$m'=0$) tile, have opposite winding numbers due to the complex conjugation in \cref{eq:Hfi_symmetry}.

\begin{figure*}
\centering
    
     \includegraphics[width=\textwidth]{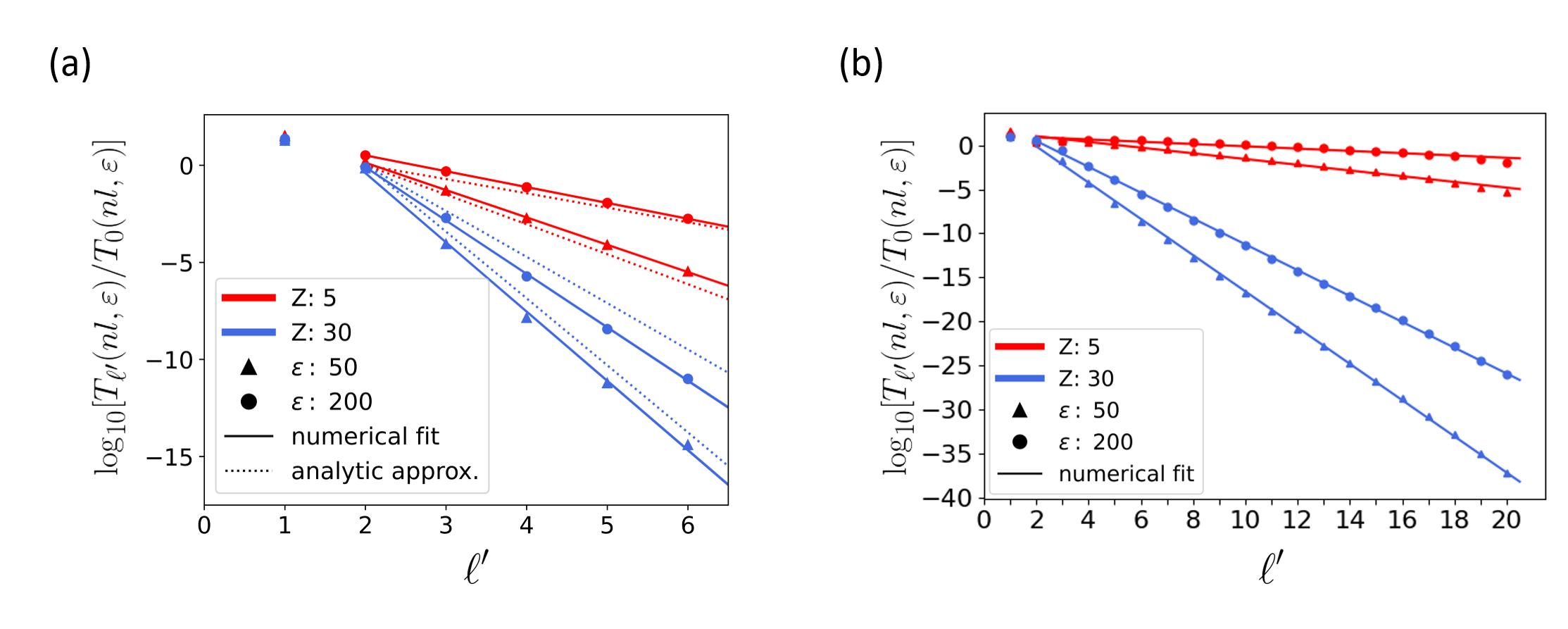}
     
     \hfill
    \caption{Sequence in \cref{eq:P_falloff} (normalised by the $\ell'=0$ term) plotted for boron and zinc for ionization energies 50 eV and 200 eV above the (a) K  and (b) $\text{L}_1$ edge. Solid lines are a guide to the eye, emphasising the linear trend. Dotted lines show the prediction of the approximate analytic expression, \cref{eq:analytic_approx_falloff}, for K shell ionization. }
    \label{fig:P_falloff}
\end{figure*}

The transition potentials in each tile in \cref{fig:Hn0_plots} are normalised to a common maximum magnitude to make the phase structure clear. However, this obscures that the magnitudes vary significantly. In particular, they tend to decrease with increasing $\ell'$. Transitions which are unlikely to be realized can be neglected with little consequence on the accuracy of the calculation. Indeed, simulation relies on being able to reduce the in-principle infinite number of final states to sum over in \cref{eq:inel_diff_patt} ($\ell'$ being any non-negative integer) to a limited number that capture the majority of the inelastic scattering. The 
transition to different final states $f$ is, for the same incident wavefield, bounded\footnote{This follows by applying the triangle inequality to the integrated intensity on both sides of \cref{eq:Y_eqns_soln}, an idea developed further in the Supplementary Material.} by the integrated intensity in $H_{fi}$. We therefore consider the rate of convergence of the following sequence in $\ell'$:
\begin{align}
  T_{\ell'}(n\ell,\varepsilon) = \sum_{m,m'} \int d\textbf{r}_\perp \, | H_{n\ell m \rightarrow \varepsilon\ell' m'}(\textbf{r}_\perp)|^2 \;.
    \label{eq:P_falloff}
\end{align}
A remarkable pattern emerges when the sequence (normalised by the $\ell'=0$ term) is plotted on a log scale as illustrated in \cref{fig:P_falloff}. Beginning from $\ell'= 2$ for K shell and $\ell'=4$ for $L_1$ shell, the terms of the sequence fall off exponentially (linear behaviour with the vertical scale plotted logarithmically). This fall-off varies depending on several experimental parameters: the atomic number $Z$ is most significant, followed by the energy loss above ionization threshold $\varepsilon$. The dependence on accelerating voltage is not shown because it is very weak, having largely cancelled out in the normalisation by the $\ell'=0$ term. As shown in the Supplementary Material, the dominant energy dependence of the inelastic transition potentials comes through the approximate scaling with $1/k_z^3$: decreasing the beam energy would decrease $k_z$ and thus increase $T_{\ell'}(n\ell,\varepsilon)$.

As derived in the Supplementary Material, an approximate analytic expression for the fall-off observed for the K shell is given by

\begin{align}
    \log_{10}\left[\frac{T_{\ell'}(n\ell,\varepsilon)}{T_{0}(n\ell,\varepsilon)}\right] \sim  2\ell' \left[\log_{10} e + \frac{1}{2}\log_{10} |\varepsilon/R_d - Z^{2/7}| - \log_{10}(\zeta)\right] + \text{terms weakly depending on} \, \ell' \;,
    \label{eq:analytic_approx_falloff}
\end{align}
where $\zeta \approx Z$ and $R_d=13.6$ eV. A correction factor of $Z^{2/7}$ is included based on empirical considerations. This approximation is plotted in \cref{fig:P_falloff}(a) by dotted lines (though not \cref{fig:P_falloff}(b) since the analytic approximations we have used do not hold for $L_1$ edge), and seen to be in good qualitative agreement with the more detailed calculation. \Cref{eq:analytic_approx_falloff} shows specifically that $Z$ and $\varepsilon$ have approximately equal effects on the fall-off, albeit in opposite directions. For instance, doubling $Z$ and halving $\varepsilon$ leaves the fall-off approximately unchanged. 

\section{Case study – monolayer materials}

To clarify the similarities and differences, let us first re-derive the elastic DPC expression before proceeding to EFDPC. For an elastic wavefield $\Psi$, the first moment of the diffraction intensity is given by
\begin{align}
    {\textbf C} = \int {\bf k}_\perp \left| \Psi({\bf k}_\perp) \right|^2 d{\bf k}_\perp \; = \frac{i}{2\pi} \int \psi({\bf r}_\perp) \left[\nabla_{{\bf r}_\perp} \psi^*({\bf r}_\perp)\right] d{\bf r}_\perp \;,
    \label{eq:S}
\end{align}
where $\psi({\bf r}_\perp)$ is the exit surface wave function with Fourier transform ${\mathcal F}\{\psi\}=\Psi({\bf k}_\perp)$. The second equality follows by the Fourier derivative theorem. Within the phase object approximation, the exit surface wave function resulting from elastic scattering takes the form
\begin{align}
    \psi(\textbf r_\perp) = e^{i \phi(\textbf r_\perp)} \psi_0(\textbf r_\perp - \textbf{R}) \;,
    \label{eq:elastic_phase_obj}
\end{align}
where the transmission function phase $\phi({\bf r}_\perp)=\sigma V({\bf r}_\perp)$ (i.e. is proportional to the specimen projected potential $V({\bf r}_\perp)$), $\psi_0$ is the entrance surface wave field, and $\textbf{R}$ is the probe position on the specimen surface. Substituting \cref{eq:elastic_phase_obj} into \cref{eq:S} yields
\begin{align}
    \textbf C({\bf R}) = \frac{1}{2\pi} \int d{\bf r}_\perp [\nabla_{\textbf{r}_\perp} \phi(\textbf{r}_\perp)] |\psi_0(\textbf{r}_\perp - \textbf{R})|^2 + \frac{i}{2\pi} \int d{\bf r}_\perp \psi_0(\textbf{r}_\perp - \textbf{R}) \nabla_{\textbf{r}_\perp}\psi^*_0(\textbf{r}_\perp - \textbf{R}) \;.
    \label{eq:elastic_com}
\end{align}
When the probe-forming aperture is symmetric, the second term in \cref{eq:elastic_com} is identically zero, and thus \cite{lazic_phase_2016}
\begin{align}
    \textbf{C}({\bf R}) = \frac{1}{2\pi} |\psi_0(\textbf{R})|^2 \star \nabla \phi(\textbf{R}) \;,
    \label{eq:elastic_com_reduced}
\end{align}
where $\star$ denotes cross correlation. 

In the inelastic regime, for a single transition the exit surface wavefield within the multiplicative-object approximation is given by both the transition of an elastic to an inelastic wavefield as per \cref{eq:Y_eqns_soln} and transmission through the specimen as per \cref{eq:elastic_phase_obj}:
\begin{align}
    \psi(\textbf{r}_\perp) = -i \sigma H_{fi}(\textbf{r}_\perp) e^{i\phi(\textbf{r}_\perp)} \psi_0(\textbf{r}_\perp-\textbf{R}) \;.
    \label{eq:psi_exit}
\end{align}
To describe the inelastic case for inner shell ionization, not only must we sum over all final states as per \cref{eq:inel_diff_patt} but it is also appropriate to sum over the different possible initial states which produce the same energy loss. Specifically, neglecting fine structure, this means summing over initial magnetic quantum number $m$ and spin states $s$. The appropriate generalisation of \cref{eq:inel_diff_patt} is then
\begin{align}
    \mathcal{I}({\bf k}_\perp) = \sum_{\substack{\text{atoms} \\ m,s,\, \ell',\, m'}}  \left| \Psi_{n\ell m \rightarrow \varepsilon \ell' m'}({\bf k}_\perp) \right|^2 \;,
    \label{eq:Psi_sum}
\end{align}
where $n$ and $\ell$ are fixed by the EELS edge chosen and $\varepsilon$ is fixed by the excitation energy above threshold. We will use the shorthand notation $\sum_{a, \, fi}$ for this summation in subsequent equations, with $a$ distinguished from $f$ to emphasise that we must sum over multiple atomic sites as well as final states for each atom.

Replacing the coherent intensity in \cref{eq:S} with the incoherent intensity in \cref{eq:Psi_sum} using the wavefields of \cref{eq:psi_exit} yields
\begin{align}
   {\bf C} (\textbf{R}) &= \frac{\sigma^2}{2\pi}  |\psi_0({\bf R})|^2 \star \left[\sum_{a, \, fi} |H_{fi}({\bf R})|^2 \nabla \phi({\bf R})\right] \nonumber \\ & \qquad + \frac{i \sigma^2}{2\pi} \sum_{a, \, fi} \int H_{fi}(\textbf{r}_\perp) \psi_0(\textbf{r}_\perp-\textbf{R}) \nabla_{\textbf{r}_\perp}(H_{fi}(\textbf{r}_\perp) \psi_0(\textbf{r}_\perp-\textbf{R}))^*  \, d \textbf{r}_\perp \;.
    \label{eq:com}
\end{align} 
Akin to elastic DPC, it can be shown (see Appendix A) that the second term in \cref{eq:com} is identically zero for a rotationally-symmetric, aberration-free probe when initial and final states (specifically, $m$ and $m'$) are summed over. Even when the probe contains aberrations which break the rotational symmetry, the second term in \cref{eq:com} is at least an order of magnitude smaller than the first term. We thus neglect the second term to yield
\begin{align}
   {\bf C}(\textbf{R}) = \frac{\sigma^2}{2\pi}  |\psi_0({\bf R})|^2 \star \left[\sum_{a, \, fi} |H_{fi}({\bf R})|^2 \nabla \phi({\bf R})\right] \;.
    \label{eq:com_reduced}
\end{align} 
This inelastic DPC expression is very similar in form to the elastic CoM expression in \cref{eq:elastic_com_reduced}, save now that the phase gradient is modulated by the magnitude squared of the inelastic transition potentials. In particular, the contrast does not arise from the inelastic scattering (interpreted to be the $H_{fi}({\bf R})$ transition potential) but rather from the elastic scattering (interpreted to be $\phi({\bf R})$, the phase of the elastic scattering transmission function). The result can thus be considered an example of preservation of elastic contrast \cite{howie1963inelastic,brown_addressing_2016}, where contrast deriving from the elastic scattering is retained in the inelastic signal.

Strictly speaking, \cref{eq:com_reduced} means the DPC image is no longer an exact gradient. However, if the inelastic transition potential is approximately constant in the vicinity of the atom sites then
\begin{align}
    \sum_{a, \, fi} |H_{fi}|^2 \nabla \phi \approx \nabla \left( \sum_{a, \, fi} |H_{fi}|^2  \phi \right) \;.
    \label{eq:gradient_approx}
\end{align}
When this approximation holds, the EFDPC signals can be integrated in the same way as DPC signals to give an integrated DPC (iDPC) signal \cite{Kottler:07}:
\begin{align}
    \left(\phi \sum_{a,\, fi} |H_{fi}|^2\right)(\textbf{R}) \approx  \mathcal{F}^{-1}\left\{\frac{\mathcal{F}\left\{ C_x \right\}(\textbf{k}_\perp) + i \mathcal{F}\left\{ C_y \right\}(\textbf{k}_\perp)}{2\pi i (k_x + i k_y)\mathcal{F}\left\{ |\psi_0|^2 \right\}(\textbf{k}_\perp)} \right\}(\textbf{R}) \;,
    \label{eq:EF_phase}
\end{align}
where $\textbf{k}_\perp = (k_x, k_y)$ are taken to be Fourier space coordinates and $C_x$ and $C_y$ are the components of \cref{eq:com_reduced}.

\begin{figure*}
\centering
    \includegraphics[width=\textwidth]{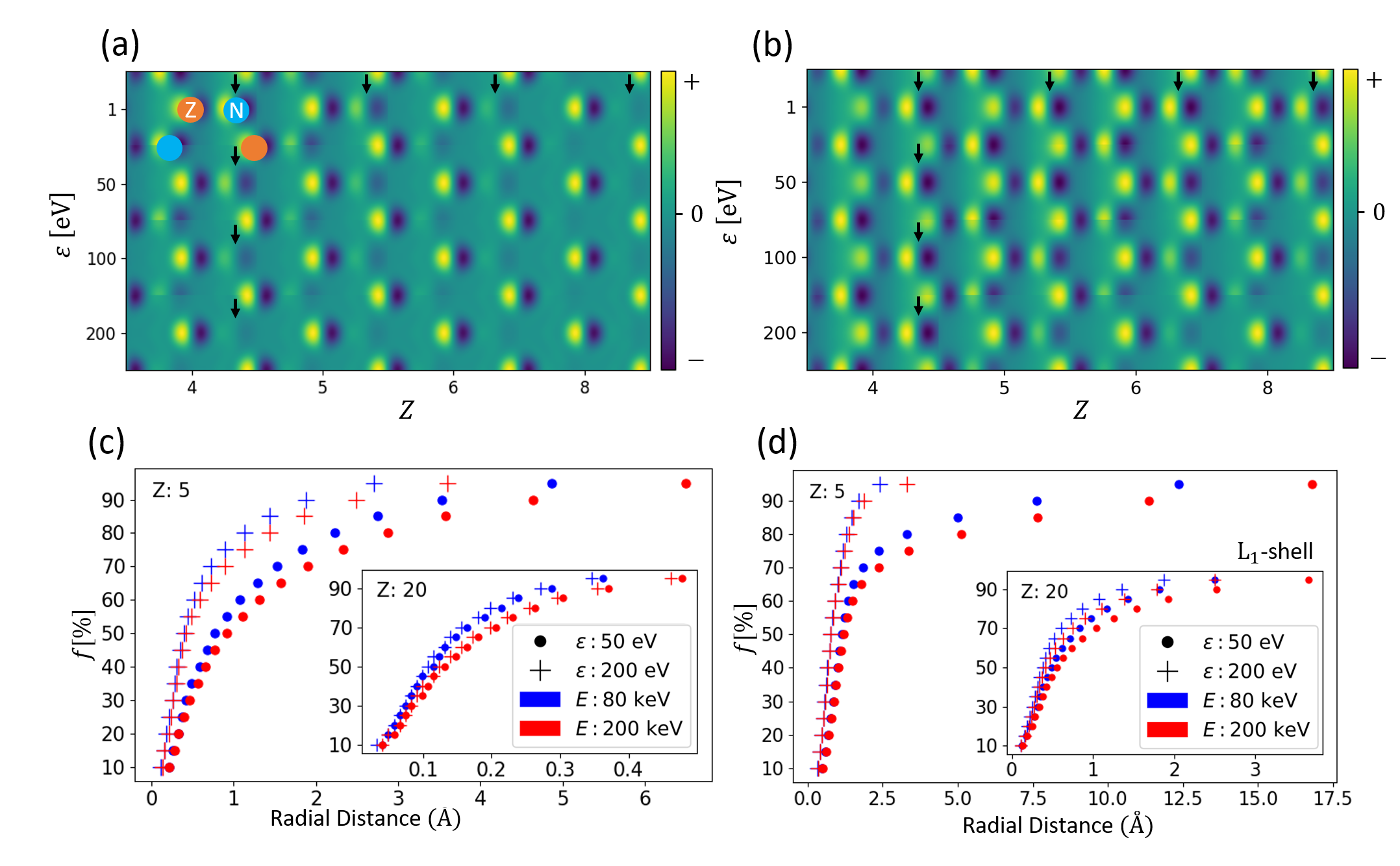}
    \caption{(a) K shell and (b) $\text{L}_1$ shell  EFDPC of the first moment along the $x$-direction for a detector spanning around $\pm$100 mrad in the $x$- and $y$-directions tableaus assuming a hBN-like structure and varying $Z$, the atomic number of the atoms at what in hBN would be the boron site, and $\varepsilon$. These simulations assume 300 keV beam electrons and a 15.7 mrad probe-forming convergence semiangle. The arrows indicate nitrogen sites where there is a gradual reduction in contrast with increasing $Z$ (of the atom being ionized) and/or energy above threshold. Each tile in the tableaus is displayed with the full colour scale spanning its individual minimum to maximum values. To convey in more detail the spatial localization of the transition potentials, we show what percentage of the integrated magnitude squared of the transition potential is contained within a given radius (see  \cref{eq:localization_expr}) for  (c) K shell (d) $\text{L}_1$ shell for all possible combinations of $Z:5, 20$, $\varepsilon: 50,200$ eV and beam energy $E: 80, 200$ keV up to (a,c) $\ell'=5$ and (b,d) $\ell'=15$ inclusive.}
    \label{fig:localization}
\end{figure*}

The interpretation of \cref{eq:com_reduced,eq:EF_phase} depends on the spatial extent of the magnitude squared of the transition potentials, $|H_{fi}|^2$. If they are so delocalised that $\sum_{a, \, fi} |H_{fi}|^2$ would be approximately constant throughout the material then the inelastic DPC signal would be proportional to the elastic DPC signal, with contrast from all atoms visible. (This is consistent with the results of Beyer et al. \cite{beyer2020influence} for plasmon-loss-filtered, atomic-resolution DPC.) If they are sufficiently localised as to become negligible over the interatomic distance in the direction transverse to the beam axis then only the columns containing the ionized elements will appear in the image, producing element selective contrast.\footnote{Eq. (\ref{eq:gradient_approx}) would only give an element specific exact derivative if $\sum_{a, \, fi} |H_{fi}|^2$ was a top-hat function encompassing all of the elastic potential of the atom of interest and none of the elastic potential of neighbouring atoms. This is not the case, and so Eqs. (\ref{eq:gradient_approx}) and (\ref{eq:EF_phase}) are only approximations. Despite this, Figs. \ref{fig:localization}(a) and (b) clearly show contrast qualitatively consistent with differential phase contrast. This is somewhat reminiscent of the distinction between first moment imaging using a pixel detector and differential phase contrast using a segmented detector: the latter is only an approximation to the former, but suffices for qualitative, and often for semi-quantitative, analysis.} If they are between these limits, being non-zero but reduced in magnitude at adjacent atomic sites, then those sites will appear in the EFDPC image with reduced contrast (relative to the elastic case).

The spatial extent of $|H_{fi}|^2$ depends on accelerating voltage of the beam, the ionization edge (i.e. the atomic species and the initial state), and the energy loss above the atom’s ionization threshold, $\varepsilon$. This is explored in the EFDPC image tableaus over $\varepsilon$ and atomic number $Z$ in \cref{fig:localization}(a) and (b). In the case Z=5 the structure is monolayer hBN; for other Z we assume a fictitious structure in which the boron atoms are replaced by atoms of that atomic number. In \cref{fig:localization}(a), the K shell ionization of that atom is assumed; in \cref{fig:localization}(b), L$_1$ shell ionization is assumed. The transition potentials become more localized with increasing atomic number and/or energy above threshold (though for the lowest energy transition shown, Be L$_1$, the delocalisation remains significant even at $\varepsilon=200$ eV), and consequently the contrast at the nitrogen atom sites, indicated by arrows, become fainter.

Hence, for element-selective contrast corresponding to a low atomic number ($\lesssim 8$) the energy window needs to be several tens of electron volts above the ionization edge. For larger atomic numbers, the energy window can be moved closer to the ionization edge. Since fewer inelastic events will occur further from the ionization edge, lower atomic numbers present the greatest challenge experimentally. The experimental images of Haas and Koch \cite{haas_momentum-_2022} of boron K shell ionization do not show boron-selective contrast, instead having contrast evident at the nitrogen sites too. This is expected since for element-selective contrast for boron the energy window need to be placed around 50 eV above the K edge. However, since this distance from the edge will yield much lower signal-to-noise, the energy filter for that experiment was likely positioned a few electron volts beyond the boron K edge.

We can quantify the spatial localization of the transition potentials by considering the radial distance from the atom within which some fraction of all transitions are expected to occur. That is, we solve
\begin{align}
    \frac{\int_0^r  \int_0^{2\pi} dr \, r \, d\theta \, \sum_{fi} |H_{fi}(r, \theta)|^2}{\int_0^\infty  \int_0^{2\pi} dr \, r \, d\theta \, \, \sum_{fi}|H_{fi}(r, \theta)|^2} = f
    \label{eq:localization_expr}
\end{align}
for $r$ for various $0< f < 1$. Since the final CoM image will contain contributions from all allowed intermediate states, \cref{eq:localization_expr} does not distinguish the individual transition contributions to the overall fraction of all expected transitions, hence the summation over final states. \Cref{fig:localization}(c) and (d) contain the solution of \cref{eq:localization_expr} for all combinations of the accelerating voltage $E: 80, 200$ keV, the ionization energy above threshold $\varepsilon: 50, 200$ eV, the atomic number $Z: 5, 20$ and the K, L$_1$ shells. The summation in \cref{eq:localization_expr} is truncated at $\ell'=5$ inclusive for \cref{fig:localization}(c) and $\ell'=15$ inclusive for \cref{fig:localization}(d) since, as implied by \cref{fig:P_falloff}, the remaining transitions are heavily suppressed. For small atomic numbers both beam energy and $\varepsilon$ are significant in the potential's localization albeit $\varepsilon$ is dominant. However, this dependence gradually flips for larger atomic numbers. The same is true, but more pronounced, for higher shells.

\section{Case study – thick specimen}

The previous section showed that if the transition potentials were sufficiently localised then EFDPC images showed differential phase contrast arising from the gradient of the elastic potential with element-selectivity resulting from the inelastic scattering event selected. This followed from the phase object approximation of \cref{eq:psi_exit}. At atomic resolution, the phase object approximation has been shown to break down for thicknesses beyond a few nanometers for elastic differential phase contrast \cite{close_towards_2015,winkler2020direct}. In this section we explore the interpretability of EFDPC images from thicker samples as the phase object approximation breaks down and dynamical diffraction becomes significant. 

\begin{figure}
    \centering
    \includegraphics[width=.6\textwidth]{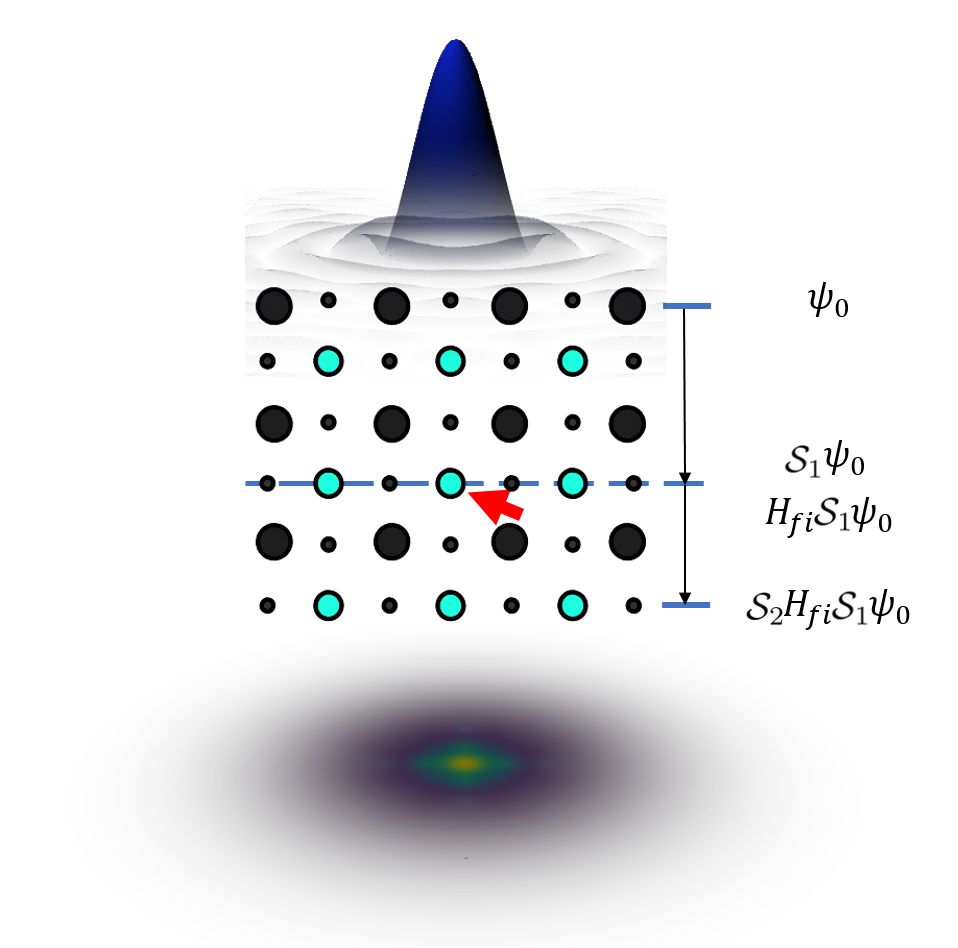}
    \caption{Schematic of the inelastic scattering process in a thick specimen of condensed matter. The incident wavefield $\psi_0$ is elastically propagated using the scattering matrix ${\mathcal S}_1$ to the depth of the atom being ionized, produces an inelastic wave depending on the transition potential $H_{fi}$, and is then elastically propagated using the scattering matrix ${\mathcal S}_2$ to the exit surface. This procedure is repeated for each final state of each atom of the ionized species, and the resultant diffraction patterns are summed incoherently.}
    \label{fig:calc_schem}
\end{figure}

To simulate core-loss-filtered diffraction patterns for thick samples, we use the efficient scattering-matrix-based approach of Brown et al. \cite{brown_linear-scaling_2019}, building on the transition potential relation in \cref{eq:Y_eqns_soln} by incorporating the elastic scattering of the wavefield in the sample before and after each inelastic transition. Consider a particular transition $H_{fi}$ of the particular atom indicated by the arrow in \cref{fig:calc_schem}. The elastic scattering of the entrance surface wavefield $\psi_0$ to the plane containing the particular atom in question is denoted here by the scattering matrix operator ${\mathcal S}_1$, and subsequent elastic scattering of the inelastic wavefield to the exit surface is denoted by the scattering matrix operator ${\mathcal S}_2$. Thus \cref{eq:Y_eqns_soln} generalises to a thick sample as
 \begin{align}
     \psi_{f}(\textbf r_\perp) = -i \sigma {\mathcal S}_2 H_{fi}(\textbf r_\perp){\mathcal S}_1 \psi_i(\textbf r_\perp) \;,
     \label{eq:Y_eqns_soln_thick}
 \end{align}
where the scattering matrix operators are understood to effect elastic scattering through the crystal. As before, the total inelastic diffraction pattern is the incoherent sum of the contributions, i.e. the intensity of the Fourier transform of \cref{eq:Y_eqns_soln_thick}, for each final state (or at least enough to obtain a converged calculation, as per the discussion of \cref{fig:P_falloff}), repeated for each atom of the same species. We evaluate the scattering matrix operator via absorptive multislice calculations. As shown by Brown et al. \cite{brown_linear-scaling_2019}, the advantage of the scattering matrix approach is that for each depth the scattering matrix can be evaluated just once but then applied to all the inelastic transitions that take place at that depth. Note that we only consider single inelastic scattering, since the probability of ionization is sufficiently small that the probability of the same beam electron ionizing multiple atoms is negligible for typical sample thickness in atomic-resolution STEM.

Taking the prototypical example of $\text{SrTiO}_3$, following Ref. \cite{close_towards_2015} we can explore the breakdown of the phase object approximation using a thickness-defocus tableau, a mosaic of DPC images across various thicknesses and probe defocus values. Because simulating for a range of thickness and defocus values is time-consuming, in the calculations that follow we have restricted the extent of the detector to 35 mrad, which in the methodology of Ref. \cite{brown_linear-scaling_2019} limits the size of the ${\mathcal S}_2$ operator and thereby the computational complexity. For core-loss transitions, the contributions for higher scattering angles are more significant than in the elastic case \cite{muller2017measurement} and so limiting the detector extent results in an underestimate of the inelastic DPC contrast. However, the qualitative appearance of the images is not appreciably affected, and the detector extent would anyway be limited in practice. We also limit the final states to $\ell' \le 2$, since, as per \cref{fig:P_falloff}, the terms $\ell' \geq 3$ are small enough to be neglected.

\begin{figure*}
\centering
         \centering
         \includegraphics[width=.4\textwidth]{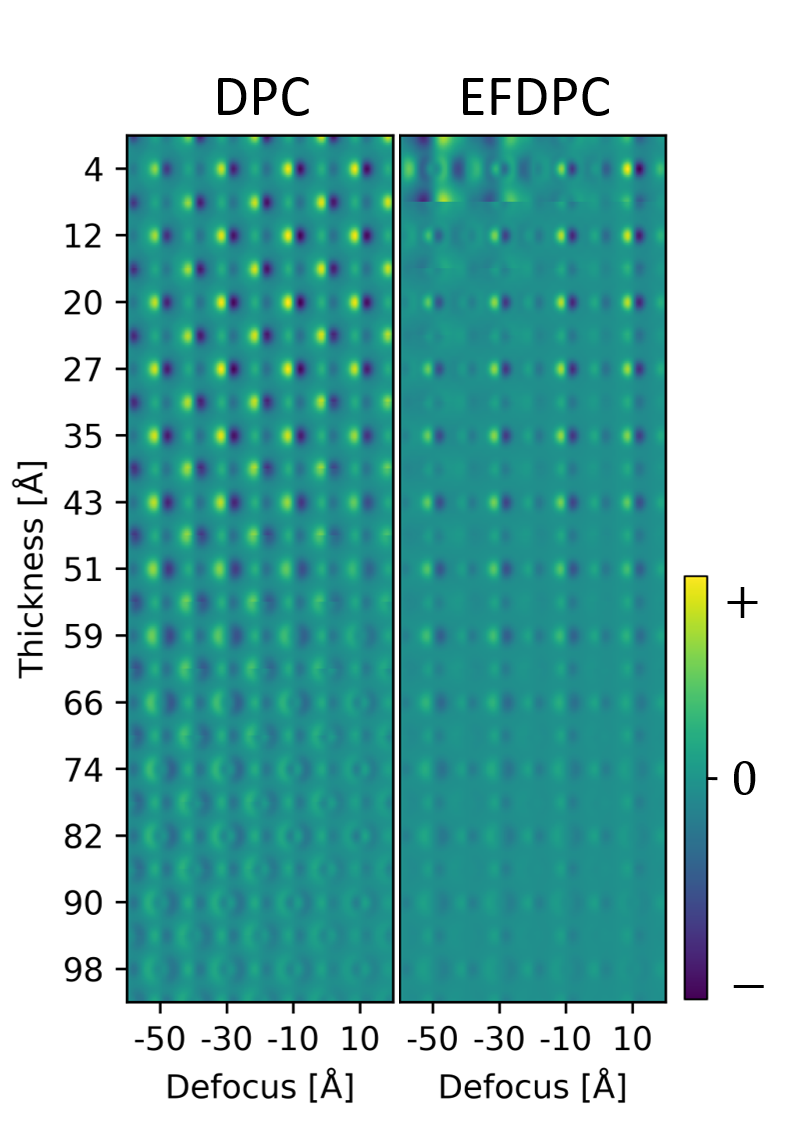}
         \caption{Thickness-defocus tableaus of $\text{SrTiO}_3$  elastic first moment ($x$-direction) DPC images (left) and first moment ($x$-direction) EFDPC 10 eV beyond the titanium L$_1$ edge (right), where underfocus is negative. All images assume a 15.7 mrad probe-forming aperture, a detector extending out to 35 mrad scattering angle and 300 keV beam energy. Within each tableau, all tiles are displayed on the same colour scale such that relative contrast can be compared between tiles. With elastic DPC normalized by thickness and EFDPC by the thickness squared here, the contrast within each unit cell should be independent of thickness within the phase-object approximation's domain of validity. This contrast is seen to decrease beyond about 35 {\AA} for both regimes, evidence of the phase-object approximation starting to break down.}
         \label{fig:EFSTEM_tab_normalized}
\end{figure*}

To make evident when the phase object approximation breaks down quantitatively, we normalise each DPC image in the tableau according to the sample thickness. \Cref{eq:elastic_com_reduced} shows that the elastic DPC image will increase linearly with thickness (assuming a specimen periodic along the beam direction) since the phase of the transmission function $\phi$, which is proportional to the projected potential, is linearly proportional to thickness. Hence, normalising the elastic DPC images by the sample thickness will lead to the contrast within each unit cell being independent of thickness within the phase object approximation's domain of applicability. This is seen in the elastic DPC tableau in the left panel in \cref{fig:EFSTEM_tab_normalized}: up to a thickness of roughly 35 {\AA}, the contrast is largely independent of thickness. Beyond that thickness, the decrease in contrast is evidence of the quantitative breakdown of the phase object approximation. For still larger thicknesses, the pattern starts to change as the breakdown of the phase object approximation impacts the interpretability of the images. For the present parameters the depth of focus is around 130 {\AA} and so the defocus dependence in the figure is weak over the range shown, though the finding that in the phase object approximation the contrast is maximised when the probe is focused on the specimen midplane \cite{close_towards_2015} is perceptible in the images at larger thicknesses being more interpretable for greater defocus into the sample.

In the inelastic case, \cref{eq:com_reduced}, in addition to the linear dependence of $\phi$ on sample thickness, the incoherent summation of contributions from the atoms along the column is also linear in thickness. Hence, normalising the EFDPC images by the square of the sample thickness will lead to the contrast within each unit cell being independent of thickness within the phase object approximation's domain of applicability. This is seen in the EFDPC tableau for the titanium L$_1$ edge in the right panel in \cref{fig:EFSTEM_tab_normalized}. The quantitative domain of validity of the phase object approximation is seen to be consistent with the elastic case, with very similar contrast seen for thicknesses below 35 {\AA} but a decrease in contrast at larger thicknesses. However, excepting for large defocus in the thinnest sample, the pattern in the EFDPC image persists to larger thicknesses than in the elastic case.

\begin{figure*}
\centering

    \includegraphics[width=1\textwidth]{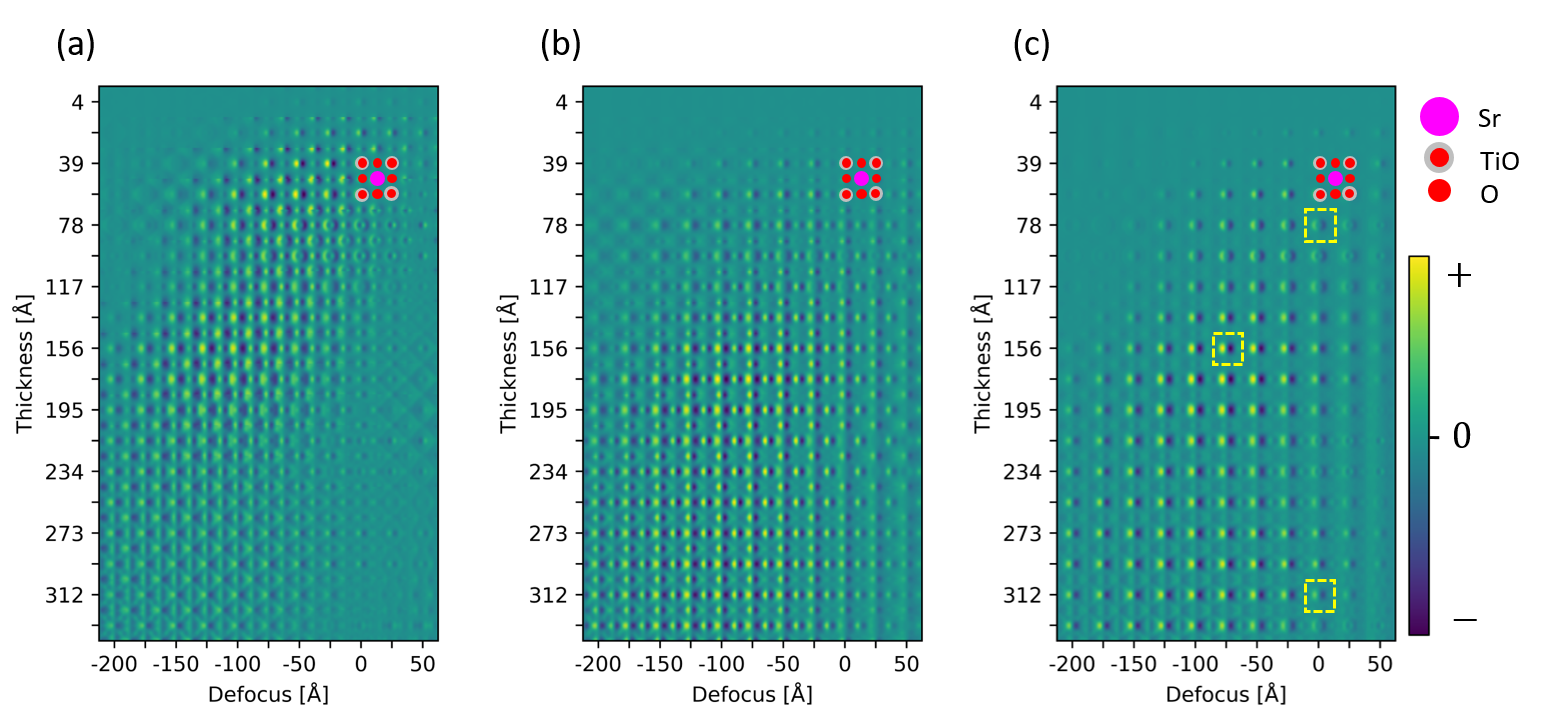}
     \hfill
    \caption{Thickness-defocus tableaus of $\text{SrTiO}_3$ for (a) elastic first moment ($x$-direction) DPC images, (b) first moment ($x$-direction) EFDPC images 10 eV beyond the oxygen K edge, and (c) first moment ($x$-direction) EFDPC images 10 eV beyond the titanium L$_1$ edge. All images assume a detector extending out to 35 mrad scattering angle, a 15.7 mrad probe-forming aperture and 300 keV beam energy, where underfocus is negative. Within each tableau, all tiles are displayed on the same colour scale such that relative contrast can be compared between tiles. Elastic DPC remains interpretable for specimens of thickness of around 40 {\AA}  given the probe is focused into the mid-plane of the specimen. EFDPC remains interpretable throughout the entire range observed. The emphasised tiles in (c) correspond to the parameters used in \cref{fig:per_layer}.}
    \label{fig:EFSTEM_tab_unnormalized}
\end{figure*}

To show that this qualitative robustness persists to still larger thicknesses, \cref{fig:EFSTEM_tab_unnormalized} shows thickness-defocus tableaus for (a) elastic DPC, and for EFDPC of (b) the oxygen K edge, and (c) titanium L$_1$ edge. These extended tableaus are not thickness-normalised and more clearly show that a defocus value near the specimen midplane tends to maximise the contrast. In contrast to the elastic case, where we see that the defocus-thickness combinations over which the images show clear differential phase contrast is somewhat limited, in the inelastic case the images show clear differential phase contrast over a wide range of thickness and defocus values. Atom-selective contrast is evident, with only the Ti sites clearly visible in the titanium L$_1$ edge case. In the oxygen K edge case, we see both the pure O and TiO columns, with the latter showing higher contrast due to the larger gradient of the elastic potential on that more-strongly-scattering TiO column. Inelastic DPC thus constitutes an interesting compromise: it contains phase-contrast-like elements through preservation of elastic contrast, but maintains interpretable contrast over a wide thickness as is more usually found in incoherent imaging (somewhat reminiscent of energy filtered TEM \cite{brown_addressing_2016}).

\begin{figure*}
\centering
    \includegraphics[width=\textwidth]{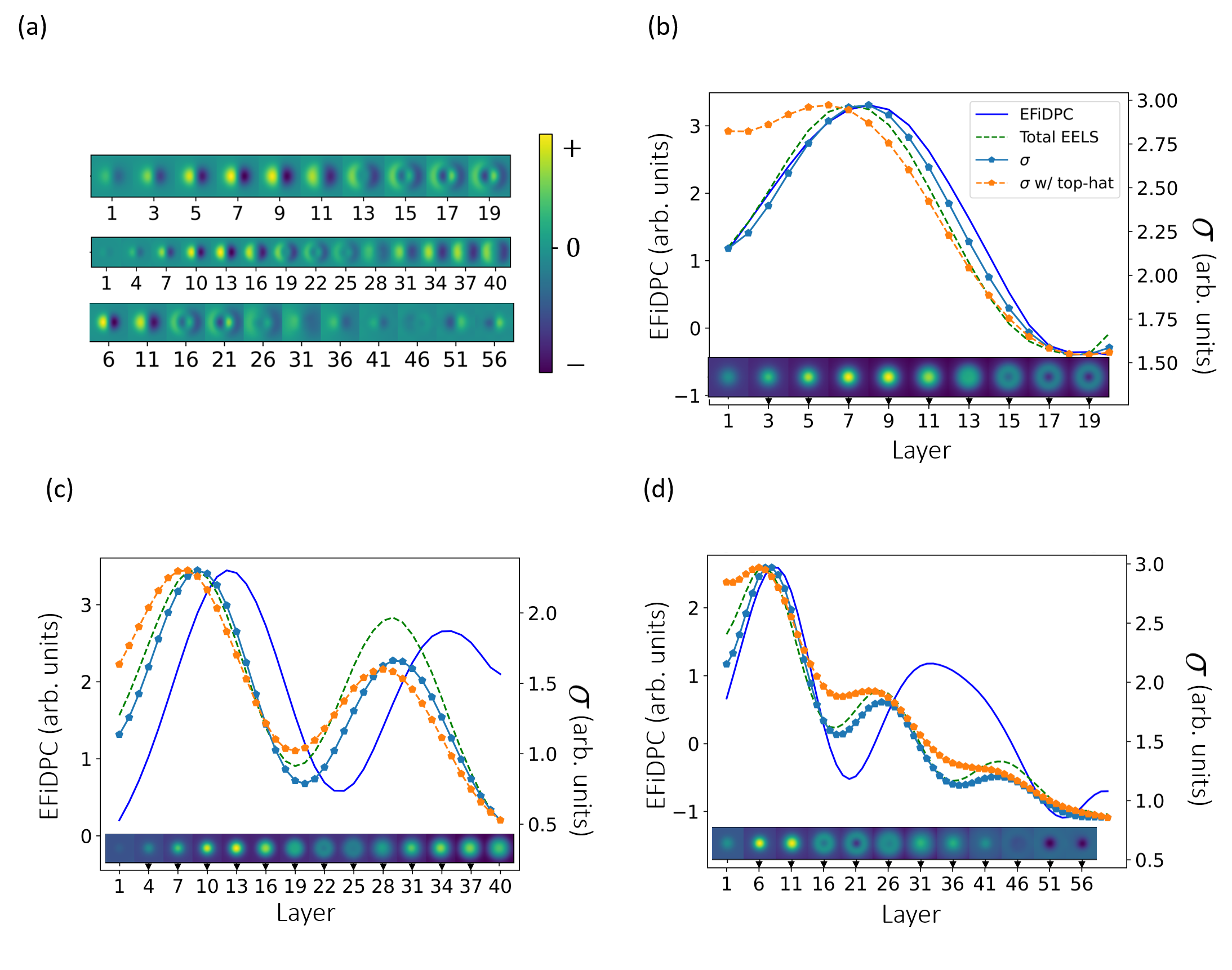}
     \hfill
    \caption{(a) Layer contributions (from a sub-set of layers) to the first moment ($x$-direction) EFDPC images for the three different thickness and defocus parameter combinations indicated by the coloured boxes in \cref{fig:EFSTEM_tab_unnormalized}(c), specifically 78.1 {\AA} (20 layers) and surface defocus (top), 156.2 {\AA} (40 layers) and $-75$ {\AA} defocus (middle), and 234.3 {\AA} (60 layers) and surface defocus (bottom), all assuming $\text{SrTiO}_3$ specimen, 300 keV beam electrons, 15.7 mrad aperture and energy filter placed 10 eV beyond the Ti $\text{L}_1$ edge. A layer is taken to be the depth of a unitcell i.e. 3.9 {\AA}. For each of the three cases respectively, (b), (c) and (d) plot the energy-filtered, integrated DPC (EFiDPC), integrated inelastic intensity (total EELS), scattering cross-section ($\sigma$, \cref{eq:cross_section}), and the scattering cross-section with a top-hat function ($\sigma$ w/ top-hat, \cref{eq:cross_section_disk}) for the probe on the TiO column as a function of layer depth. Insets show the 2D EFiDPC maps about the TiO column. All plots are scaled such that maxima and minima align, so while the EFiDPC signal given on the left axis and the cross-section given on the right correctly conveys the zero location and the contrast, the scales are in arbitrary units. The broad similarity between the plots results from the underlying channelling of the probe being common to each signal.}
    \label{fig:per_layer}
\end{figure*}

However, this robust interpretability does not mean that channelling effects are absent. To see this, we decompose the incoherent sum over final states in \cref{eq:Psi_sum} to the total contribution from unit-cell-thick layers perpendicular to the optic axis. \Cref{fig:per_layer}(a) shows this layer decomposition for the tiles in the coloured boxes in \cref{fig:EFSTEM_tab_unnormalized}(c), these defocus-thickness parameter combinations --- $\{t=78 \; {\rm \AA}, \; \Delta f = 0 \; {\rm \AA}\}$, $\{t=156 \; {\rm \AA}, \; \Delta f = -75 \; {\rm \AA}\}$, and $\{t=234 \; {\rm \AA}, \; \Delta f = 0 \; {\rm \AA}\}$ --- being diverse choices over the parameter space shown. We see that the contribution from different layers varies. The layers which show the clearest differential phase contrast are also those which contribute most to the signal, which helps explain the robustness of the total contrast, even though the contribution from some layers is far less directly interpretable. Which layers contribute the most varies widely for the different defocus-thickness parameter combinations: in \cref{fig:per_layer}(b) the layers near the midplane dominate the intensity; in \cref{fig:per_layer}(c) the layers prior and subsequent to the mid-plane provide comparable contributions; in \cref{fig:per_layer}(d) the layers prior to the midplane dominate. Note too that the layers which contribute the most are not necessarily those at the nominal in-focus plane. Nevertheless, in each case the sum over layers produces a fairly similar total image, consistent with the fairly uniform contrast in \cref{fig:EFSTEM_tab_unnormalized}(c).

That different depths contribute differently has been well established in inelastic STEM imaging, both in the context of high angle annular dark field imaging and electron energy loss spectroscopy \cite{hillyard1993annular,dwyer2003scattering,allen2003lattice,voyles2004depth}. In both cases, the underlying cause is the way the probe electron wavefield evolves through the sample. The intensity distribution of an atomically fine electron probe on an atomic column oscillates with propagation along the columns, at some depths peaking up on the columns and at others spreading out. One way of visualising this is to evaluate the integrated elastically-scattered electron probe intensity within a certain radius of the column, i.e.
\begin{equation}
    \sigma(z) = \int_{\text{disk}} d\textbf{r}_\perp \, |\psi_0(\textbf{r}_\perp, z)|^2 
    \label{eq:cross_section_disk}
\end{equation}
(this is not conventionally a cross-section, but the reasons for this notation will become clear presently). For the probe centred on the Ti column, this is plotted as the orange line, labelled ``$\sigma$ w/ top-hat'', in \cref{fig:per_layer}(b)-(d) for the three different thickness and defocus parameter combinations considered previously, and this integrated intensity is seen to vary with layer depth. That this behaviour largely underpins the layer dependence of the EFDPC signal can be motivated by the following series of models of increasing complexity.

The cross-section expression for electron energy loss spectroscopy is given by \cite{allen2003lattice}
\begin{equation}
    \sigma_{\text{layer}}(z) = \int_{z}^{z+u_z} dz \int d\textbf{r}_\perp \, |\psi_0(\textbf{r}_\perp, z)|^2 V_{\rm eff}(\textbf{r}_\perp, z) \;,
    \label{eq:cross_section}
\end{equation}
where $u_z$ is the layer thickness, $V_{\rm eff}(\textbf{r}_\perp, z)$ is an effective scattering potential for the inelastic transition and incorporating the detector geometry\footnote{For notational simplicity, we have absorbed into $V_{\rm eff}$ the prefactors that more conventionally appear in front of \cref{eq:cross_section}.}, and this model neglects elastic scattering after the inelastic scattering event. The blue dotted lines, labeled $\sigma$, in \cref{fig:per_layer}(b)-(d) show the titanium L$_1$ edge signal calculated in this model, assuming a 35 eV energy window above threshold and a very large detector collection angle. The radius of the disk integration region used to evaluate \cref{eq:cross_section_disk} was chosen to be that containing $60\%$ of the effective scattering potential $V_{\rm eff}$ since it gave the most favourable agreement with the results of \cref{eq:cross_section}. Nevertheless, and while not identical, the similarity in the plots supports the channelling behaviour being the primary cause of the oscillations seen. 

The green lines in \cref{fig:per_layer}(b)-(d) were calculated using the transition potential model of \cref{eq:Y_eqns_soln_thick}, assuming an energy filter 10 eV above the titanium L$_1$ edge and a large detector. Since this quantity is the would-be-observed intensity on an EELS spectrum at the corresponding energy loss, the signal is labelled ``total EELS''. By including absorption due to thermal scattering after the ionization event and restricting to a single energy loss, the assumptions of this calculation differ somewhat from that of the cross-section expression. (Elastic scattering after the ionization event is also included, but that redistribution will not much affect the total inelastic intensity.) Nevertheless, the total EELS signal and the cross-section expression are in close agreement, suggesting those differences have only a small effect on the depth dependence of the signal.

Since these plots considered the probe on the column site, where the EFDPC signal is zero, the blue lines in \cref{fig:per_layer}(b)-(d) show instead the energy-filtered iDPC (EFiDPC) signal, which is peaked on the column. The EFDPC $x$ and $y$ signals are not strictly partial derivatives of the same function even in the thin sample regime, \cref{eq:com_reduced}, but the procedure on the right hand side of \cref{eq:EF_phase} can still be applied and, as shown in the select layer insets in \cref{fig:per_layer}(b)-(d), the resultant images have the expected qualitative form of an iDPC signal when the DPC signals are directly interpretable. Though the blue lines show more differences from the other lines, which we attribute largely to elastic scattering after the inelastic scattering event having some impact on the DPC value,  their general shape is broadly quite similar, with clear correspondence between the peaks and troughs. This is consistent with channelling being the underlying mechanism for the variation in contributions from different layers.

\section{Practical considerations}
\label{sec:practical}
Having focused primarily on qualitative interpretation, contrast formation mechanisms and imaging dynamics, we have thus far presented results assuming infinite dose. Though the experimental results of Haas and Koch \cite{haas_momentum-_2022} clearly show EFDPC to be achievable even from a monolayer, one challenge for this imaging mode is that the cross-section for core-loss scattering is typically several orders of magnitude lower than that of elastic scattering, meaning high doses will be required to obtain sufficient signal to be discernable above shot noise. But before considering typical doses required for EFDPC imaging, a comment on units is warranted.

We display EFDPC images simulated using the transition potential approach with arbitrary units not because the units are unknown but because they are not intuitive. There are two reasons for this. First, by considering ionization to continuum states the transition potential calculation of core-loss filtered diffraction intensities via Eq. (\ref{eq:inel_diff_patt}) has units of eV$^{-1}$, which follows from the normalisation of Eq. (6) in the Supplementary Material. Putting core-loss-filtered diffraction intensities on the same scale as the wavefield normalisation requires integration over a range of energy losses, with separate inelastic scattering calculations required at each energy loss, a complication that was unnecessary for exploring the contrast formation mechanisms. Second, the scale of the first moment evaluation can depend on the scale of the wavefield. In practice, elastic DPC is usually evaluated not by Eq. (\ref{eq:S}) but rather by the normalised form
\begin{align}
    {\textbf C}_{\rm normalised} = \frac{ \int {\bf k}_\perp \left| \Psi({\bf k}_\perp) \right|^2 d{\bf k}_\perp }{ \int \left| \Psi({\bf k}_\perp) \right|^2 d{\bf k}_\perp } \;,
    \label{eq:Snormalised}
\end{align}
eliminating any wavefunction normalisation to leave the result in units of the diffraction plane coordinates (often converted to a scattering angle). This works because the total recorded intensity (the denominator in Eq. (\ref{eq:Snormalised})) depends only weakly on probe position.\footnote{That it depends on probe position at all results from scattering outside the finite detector extent when the probe is above strongly scattering columns, and the normalisation of Eq. (\ref{eq:Snormalised}) can to some extent compensate for this.} Core-loss-filtered 4D STEM is very different: the total inelastic signal varies considerably depending on where the probe is relative to columns of the element selected. Simulations (not shown) indicate that a normalisation like Eq. (\ref{eq:Snormalised}) for EFDPC reduces the qualitative similarity with elastic DPC images. But the consequence of not performing such a normalisation is that the first moment evaluation of Eq. (\ref{eq:S}) is significantly affected by the scale of the wavefunction, preventing intuitive display of EFDPC images as maps of the scattering angle.

\begin{figure*}
\centering
\includegraphics[width=0.5\textwidth]{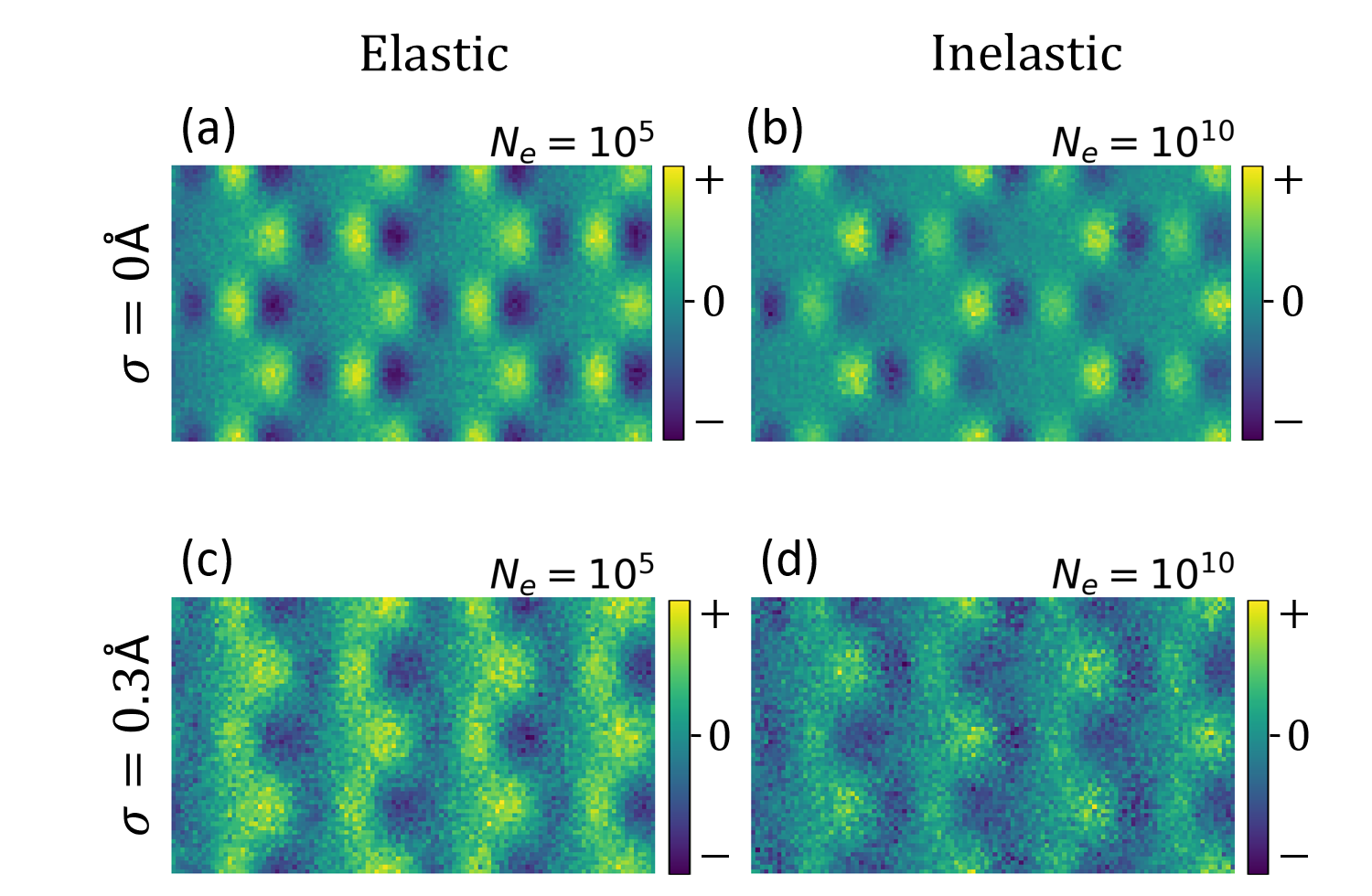}
\hfill
\caption{(a) Elastic DPC and (b) EFDPC (with a 20 eV energy window above the boron K edge) simulated images of the first moment along the $x$-direction for a detector spanning around $\pm 100$ mrad in the $x$- and $y$-directions for monolayer hBN. Calculations assume (a) $10^5$ and (b) $10^{10}$ electrons per probe position (corresponding to doses of (a) $\sim 10^{6}$ e$^{-}$/{\AA}$^2$  and (b) $\sim 10^{11}$ e$^{-}$/{\AA}$^2$). (c,d) As per (a,b) but including partial spatial incoherence assuming a Gaussian effective source distribution with standard deviation 0.3 {\AA}.}
\label{fig:hbn_noise}
\end{figure*}

Despite this, we can still explore the typical dose required for EFDPC imaging. \Cref{fig:hbn_noise} shows simulated elastic DPC (left) and boron K edge EFDPC (right, assuming a 20 eV energy window above the ionization threshold\footnote{Modelled using numerical integration over transition-potential-based simulations sampled every 2 eV in energy loss.}) images for a monolayer hBN including realistic (Poisson-distributed) shot noise based on the number of electrons reaching the 4D STEM detector. \Cref{fig:hbn_noise}(a) and (b) assumes the number of electron per probe position, $N_e$, to be (a) $10^5$ (or $\sim 10^6$ e$^{-}$/{\AA}$^2$ at the sampling shown) and (b) $10^{10}$ (or $\sim 10^{11}$ e$^{-}$/{\AA}$^2$ at the sampling shown), chosen to be the lowest values, to the nearest order of magnitude, for which signal is clearly discernible above the noise. Thus in this case the EFDPC image requires $10^5$ times more dose than elastic DPC for comparable contrast. Whereas \cref{fig:hbn_noise}(a) and (b) (indeed all simulations thus far) assumes perfect source coherence, \cref{fig:hbn_noise}(c) and (d) further accounts for partial spatial incoherence, included here assuming a Gaussian effective source distribution with a standard deviation of 0.3 {\AA} that broadens the features while approximately halving the signal. \Cref{fig:hbn_noise}(c) and (d) use the same dose as (a) and (b), respectively. The lower contrast is seen to manifest as more pronounced noise but the signal remains visible, suggesting that realistic effective source distributions have only a modest effect on the dose required.

\begin{figure*}
\centering
\includegraphics[height=0.8\textheight]{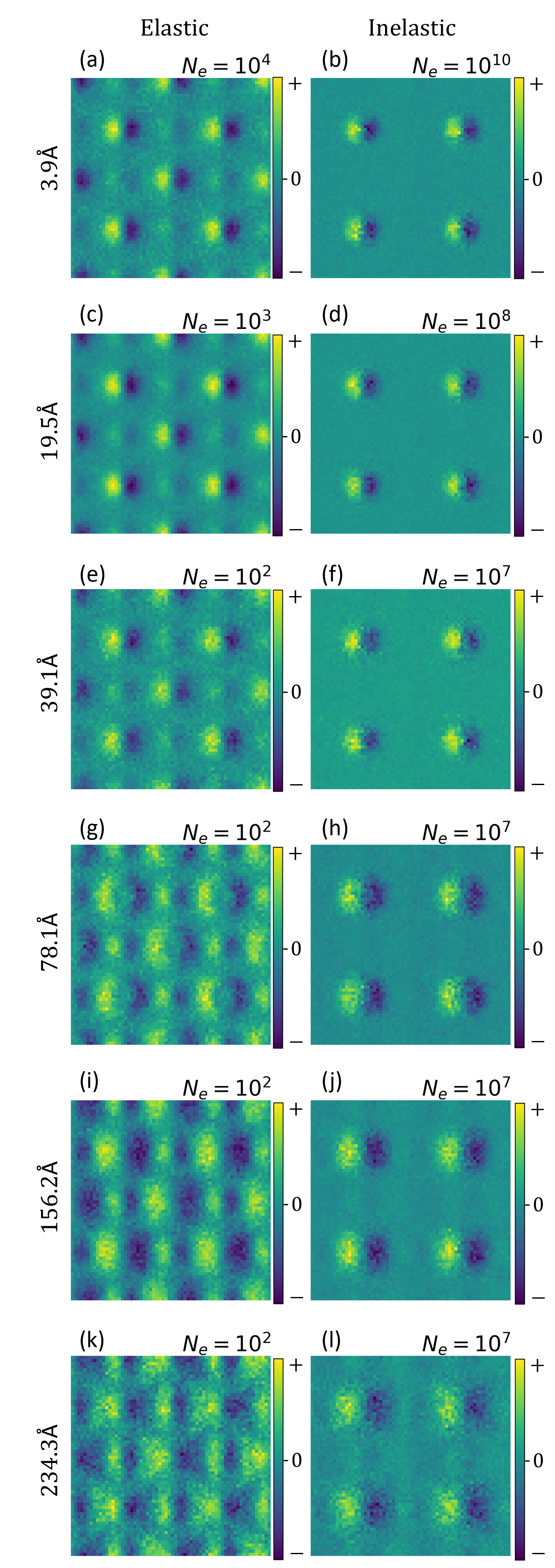}
\hfill
\caption{
Elastic DPC (left column) and EFDPC (right column, with a 20 eV energy window above the titanium L$_1$ edge) simulated images of the first moment along the $x$-direction for a detector with an angular range extending out to 35 mrad for $\text{SrTiO}_3$ for thicknesses (a,b) 3.9 {\AA} (1 layer), (c,d) 19.5 {\AA} (5 layers), (e,f) 39.1 {\AA} (10 layers), (g,h) 78.1 {\AA} (20 layers), (i,j) 156.2 {\AA} (40 layers) and (k,l) 234.3 {\AA} (60 layers). The $N_e$ values above each image correspond to the number of electrons per probe position. For example, (a) and (b) respectively assume $10^4$ and $10^{10}$ electrons per probe position (corresponding to doses of $\sim 10^5$ e$^{-}$/{\AA}$^2$ and $\sim 10^{11}$ e$^{-}$/{\AA}$^2$, respectively). All calculations assume defocus to the midplane.}
\label{fig:STO_noise}
\end{figure*}

\Cref{fig:STO_noise} compares elastic (left column) and inelastic (right column; with a 20 eV energy window above the titanium L$_1$ edge) DPC image simulations for SrTiO$_3$ at various thicknesses. (We revert to assuming assuming perfect coherence to enable direct comparison with \cref{fig:EFSTEM_tab_unnormalized}(a) and (c).) Again, the dose assumed for each case is, to the nearest order of magnitude, the lowest for which appreciable contrast is visible above the noise. \Cref{fig:STO_noise}(a) and (b) assumes a monolayer of SrTiO$_3$, which, while very hard to achieve in practice, enables comparison with the monolayer hBN results in \cref{fig:hbn_noise}(a) and (b). Cross-sections for ionizing different elements and shells will generally differ \cite{egerton2011electron}. However, as per \cref{eq:com_reduced}, EFDPC images in the phase object approximation scale with both the ionization cross-section and the (gradient of the) elastic potential. Thus while the titanium L$_1$ shell has a lower ionization cross-section than the boron K shell, Ti having a greater elastic potential than B results in the minimum dose needed for monoloayer SrTiO$_3$ being lower than that for monolayer hBN.

Subsequent rows in \cref{fig:STO_noise} show results for larger thicknesses of SrTiO$_3$. Again the dose assumed for each case is, to the nearest order of magnitude, the lowest for which appreciable contrast is visible above the noise. To a thickness of around 4 nm, increasing the thickness reduces the dose required to achieve a comparable level of noise. Beyond that, the required dose is approximately constant, though the apparent noise level increases slightly with increasing thickness. This is broadly consistent with the thickness dependence of the signal as seen in \cref{fig:EFSTEM_tab_unnormalized}(a) and (c), where the signal initially grows, then saturates and finally fades somewhat with increasing thickness. Interestingly, for the first 4 nm the relative reduction in necessary dose with increasing thickness is greater for EFDPC than for elastic DPC. This is broadly consistent with the results of \cref{fig:EFSTEM_tab_normalized}, where in the phase object approximation the elastic signal increases linearly with thickness whereas the EFDPC signal increases quadratically with thickness (due to both the accumulation of projected potential and the increase of the number of atoms ionized). We attribute the necessary dose not exactly following that scaling to the noise depending not just on the number of electrons but also on the details of the diffraction intensity distribution \cite{seki2018theoretical}. One way to further increase the signal-to-noise would be to use a larger energy window. Another, if the sample is periodic, would be repeat unit averaging.

Isolating the core-loss contribution experimentally will require reliable background subtraction. Having identified the preservation of elastic contrast as the dominant mechanism leading to EFDPC contrast from core-loss events, it seems likely that some form of DPC contrast is also present in the background contribution. The approach of Haas and Koch \cite{haas_momentum-_2022} using $\omega$-$q$ imaging has the advantage of providing spectra from which to fit the background, but data would need to be acquired sequentially at multiple slit orientations to adequately sample the momentum distribution. Conversely, the approach of core-loss-filtered 4D STEM as depicted in \cref{fig:intro} has the advantage of providing a detailed momentum distribution, but data would need to be acquired sequentially over multiple energy loss windows to effect background subtraction. Further experimental work is thus needed to determine the most effective approach to acquiring such data in practice.

\section{Conclusion}

In this work we have presented a theoretical interpretation and explored the imaging dynamics of energy-filtered DPC imaging, where the energy filter selects electrons that have caused inner shell ionization.

Within the phase object approximation applicable to thin samples, we have shown that the contrast is similar to that of elastic DPC imaging because the elastic DPC contrast is preserved within the momentum distribution of inelastically scattered electrons, over a spatial region corresponding to the range of appreciable ionization probability. For low energy losses and light elements, this means that the EFDPC signal at an energy loss corresponding to a particular element may include contrast from nearby atoms of different species, explaining why nitrogen atom sites are visible in the boron K shell EFDPC experimental results of Haas and Koch \cite{haas_momentum-_2022}. However, for heavier elements the range of the ionization interaction is much narrower, and element selective DPC contrast is expected.

For thick specimens, say beyond several nanometres, the phase object approximation breaks down quantitatively for both EFDPC and elastic DPC. However, we have shown that qualitatively the EFDPC images remain interpretable to much larger thicknesses than do elastic DPC images, a consequence of the incoherence between the contributions from the different atoms. Despite this more robust appearance, we have also shown that the contribution from different depths within the sample differ as a result of the nonuniform evolution of the probe wavefield along the atomic columns.

EFDPC thus combines the advantages of STEM EELS and elastic DPC. Indeed, the strengths of each mitigate against limitations in the other. The delocalisation of EELS is mitigated by manifesting as a modulation of the gradient of the elastic potential. This may allow the position of atoms to be more precisely determined, for instance when probing orientation relationships between a thin surface layer and a substrate. Conversely, the limited thickness range of interpretable elastic DPC images has been mitigated by the incoherence of EELS, making direct interpretation possible in much thicker samples, extending the range of samples over which reliably interpretable DPC-like contrast can be achieved. Because ionization cross-sections are relatively low, in practice a very high dose is needed for sufficient counting statistics to discern differential phase contrast from core-loss scattered electrons. However, EFDPC depending on both the number of atoms ionized and the (gradient of the) total projected potential means that the signal strength initially scales quadratically with thickness (until saturating beyond about 4 nm), leading to some enhancement of signal-to-noise ratio that makes inelastic DPC imaging more achievable for thicker samples. Further work is needed in developing procedures to acquire core-loss filtered 4D STEM data to determine whether EFDPC might be convenient and/or competitive over separately recording EELS and elastic DPC images when seeking such structural insights. Irrespective, the present results offer some initial insights into the prospects of energy-filtered 4D STEM.

\section*{CRediT authorship contribution statement}

{\bf M. Deimetry}: Methodology, Software, Formal analysis, Investigation, Writing - Original Draft, Writing - Review \& Editing, Visualization. {\bf
T.C. Petersen}: Conceptualization, Methodology, Writing - Review \& Editing, Supervision. {\bf H.G. Brown}: Methodology, Software, Writing - Review \& Editing. {\bf M. Weyland}: Conceptualization, Methodology, Writing - Review \& Editing, Supervision. {\bf S.D. Findlay}: Conceptualization, Methodology, Validation, Writing - Review \& Editing, Funding acquisition.

\section*{Declaration of competing interest}

The authors declare that they have no known competing financial interests or personal relationships that could have appeared to influence the work reported in this paper.

\section*{Acknowledgments}

This research is supported by an Australian Government Research Training Program Scholarship. This research was supported under the Discovery Projects funding scheme of the Australian Research Council (Project No. FT190100619).

\appendix
\section{Vanishing of the first moment of inelastic diffraction patterns if elastic scattering did not occur} \label{sec:elastic_pres_proof}

The diffraction intensity for wavefields $\psi_0 H_{n\ell m\rightarrow\varepsilon\ell' m'}$ summed over $m$ and $m'$ is given by
\begin{align}
   \mathcal{I}(\textbf{k}_\perp, \textbf{R}) &=  \sum_{m,m'} |(\Psi_0(\textbf{k}_\perp) e^{-2\pi i \bf k_\perp \cdot \bf R}) \otimes H_{n\ell m\rightarrow\varepsilon\ell' m'}(\textbf{k}_\perp)|^2 \nonumber \\
   & = \int d\boldsymbol{\tau} \, d\boldsymbol{\gamma} \, \Psi_0(\textbf k-\boldsymbol{\tau}) \Psi^*_0(\textbf k-\boldsymbol{\gamma}) e^{2\pi i (\boldsymbol{\tau} - \boldsymbol{\gamma})\cdot \bf R} \left[ \sum_{m,m'} H_{n\ell m\rightarrow\varepsilon\ell' m'}(\boldsymbol{\tau}) H^*_{n\ell m\rightarrow\varepsilon\ell' m'}(\boldsymbol{\gamma})\right]  \;,
    \label{eq:diff_img}
\end{align}
where $\otimes$ denotes convolution. To evaluate the term in square brackets, let us first re-write the transition matrix element as
\begin{align}
      H_{n\ell m \rightarrow \varepsilon\ell' m'}(\textbf k) = \frac{q_e^2}{4\pi^2 \varepsilon_0 k^2} \sum_{\ell''=0}^\infty \sum^{\ell''}_{m''=-\ell''} (-i)^{\ell''} Y^{m''}_{\ell''}(\hat{\textbf{k}}) \langle l' m' | \overline{l''m''}|lm\rangle R_{l',l'',l}(k)
      \label{eq:b1}
\end{align}
(which simplifies to \cref{eq:Hfi_full_exp} on using the properties of the Wigner 3$j$ symbols in \cref{eq:gaunt}). Anticipating that it will depend only on the magnitudes $\tau$ and $\gamma$ and on the dot product $\hat{\boldsymbol{\tau}}\cdot\hat{\boldsymbol{\gamma}}$, the term in square brackets can then be written as
\begin{align}
   F(\tau,\gamma,\hat{\boldsymbol{\tau}}\cdot\hat{\boldsymbol{\gamma}}) &= \sum_{m,m'} H_{n\ell m\rightarrow\varepsilon\ell' m'}(\boldsymbol{\tau}) H^*_{n\ell m\rightarrow\varepsilon\ell' m'}(\boldsymbol{\gamma}) \nonumber \\ &= 
\frac{q_e^4}{(4\pi^2 \varepsilon_0 )^2\tau^2\gamma^2} \sum_{\ell''=0}^\infty \sum_{\ell'''=0}^\infty\sum^{\ell''}_{m''=-\ell''} \sum^{\ell'''}_{m'''=-\ell'''} (-i)^{\ell''-\ell'''} Y^{m''}_{\ell''}(\hat{\boldsymbol{\tau}}) Y^{m'''*}_{\ell'''}(\hat{\boldsymbol{\gamma}}) R_{l',l'',l}(\tau) R_{l',l''',l}(\gamma) \nonumber \\ & \qquad
\sum_{m,m'} \langle l' m' | \overline{l''m''}|lm\rangle  \langle l' m' | \overline{l'''m'''}|lm\rangle  \nonumber \\ &=  \frac{q_e^4}{(4\pi^2 \varepsilon_0 )^2\tau^2\gamma^2} \frac{(2\ell'+1)(2\ell+1)}{4\pi}\sum_{\ell''=0}^\infty  R_{l',l'',l}(\tau) R_{l',l'',l}(\gamma) \begin{pmatrix}
            \ell' & \ell'' & \ell \\
            0 & 0 & 0
            \end{pmatrix}^2 \nonumber \\ & \qquad \sum^{\ell''}_{m''=-\ell''} Y^{m''}_{\ell''}(\hat{\boldsymbol{\tau}}) Y^{m''*}_{\ell''}(\hat{\boldsymbol{\gamma}})  
\nonumber \\ &=  \frac{q_e^4}{(4\pi^2 \varepsilon_0 )^2\tau^2\gamma^2} \frac{(2\ell'+1)(2\ell+1)}{4\pi}\sum_{\ell''=0}^\infty  R_{l',l'',l}(\tau) R_{l',l'',l}(\gamma) \begin{pmatrix}
            \ell' & \ell'' & \ell \\
            0 & 0 & 0
            \end{pmatrix}^2 \nonumber \\ & \qquad  \frac{(2\ell''+1)}{4\pi}  P_{\ell''}(\hat{\boldsymbol{\tau}}\cdot \hat{\boldsymbol{\gamma}}) \;,
      \label{eq:b10}
\end{align}
where the third equality follows from the orthogonality relation
\begin{align}
    (2j_3+1) \sum_{m_1,m_2} \begin{pmatrix}
j_1 & j_2 & j_3 \\
m_1 & m_2 & m_3
\end{pmatrix}
\begin{pmatrix}
j_1 & j_2 & j_3' \\
m_1 & m_2 & m_3'
\end{pmatrix}  = \delta_{j_3,j_3'} \delta_{m_3,m_3'} \;,
\end{align}
and the fourth equality follows from the addition theorem 
\begin{align}
P_{\ell''}(\hat{\textbf{x}} \cdot \hat{\textbf{y}}) = \frac{4\pi}{2\ell''+1} \sum_{m''=-\ell''}^{\ell''} Y_{\ell''}^{m''}(\hat{\textbf{y}})Y_{\ell''}^{m''*}(\hat{\textbf{x}}) \;.
    \label{eq:Y_add_th}
\end{align}
Thus we can rewrite \cref{eq:diff_img} as
\begin{align}
   \mathcal{I}(\textbf{k}_\perp, \textbf{R}) = \int d\boldsymbol{\tau} \, d\boldsymbol{\gamma} \, \Psi_0(\textbf k-\boldsymbol{\tau}) \Psi^*_0(\textbf k-\boldsymbol{\gamma}) e^{2\pi i (\boldsymbol{\tau} - \boldsymbol{\gamma})\cdot \bf R} F(\tau,\gamma,\hat{\boldsymbol{\tau}}\cdot\hat{\boldsymbol{\gamma}}) \;.
    \label{eq:diff_img2}
\end{align}

From \cref{eq:diff_img2} we may reason as follows
\begin{align}
     \mathcal{I}(-\textbf{k}_\perp, \textbf{R}) &= \int d\boldsymbol{\tau} \, d\boldsymbol{\gamma} \, \Psi^*_0(-\textbf k-\boldsymbol{\tau}) \Psi_0(-\textbf k-\boldsymbol{\gamma}) e^{2\pi i (\boldsymbol{\tau} - \boldsymbol{\gamma})\cdot \bf R} F(\tau,\gamma,\hat{\boldsymbol{\tau}}\cdot\hat{\boldsymbol{\gamma}}) \nonumber \\ &=\int d\boldsymbol{\tau} \, d\boldsymbol{\gamma} \, \Psi^*_0(-\textbf k+\boldsymbol{\tau}) \Psi_0(-\textbf k+\boldsymbol{\gamma}) e^{-2\pi i (\boldsymbol{\tau} - \boldsymbol{\gamma})\cdot \bf R} F(\tau,\gamma,\hat{\boldsymbol{\tau}}\cdot\hat{\boldsymbol{\gamma}}) \nonumber \\ &= \int d\boldsymbol{\tau} \, d\boldsymbol{\gamma} \, \Psi^*_0(-\textbf k+\boldsymbol{\gamma}) \Psi_0(-\textbf k+\boldsymbol{\tau}) e^{2\pi i (\boldsymbol{\tau} - \boldsymbol{\gamma})\cdot \bf R} F(\gamma,\tau,\hat{\boldsymbol{\tau}}\cdot\hat{\boldsymbol{\gamma}}) \nonumber \\ &= \int d\boldsymbol{\tau} \, d\boldsymbol{\gamma} \, \Psi^*_0(\textbf k-\boldsymbol{\tau}) \Psi_0(\textbf k-\boldsymbol{\gamma}) e^{2\pi i (\boldsymbol{\tau} - \boldsymbol{\gamma})\cdot \bf R} F(\tau,\gamma,\hat{\boldsymbol{\tau}}\cdot\hat{\boldsymbol{\gamma}}) \nonumber \\ &= \mathcal{I}(\textbf{k}_\perp, \textbf{R})
    \label{eq:b11}
\end{align}
where in the second line we have made the change of variables $\boldsymbol{\tau} \rightarrow -\boldsymbol{\tau}$ and $\boldsymbol{\gamma} \rightarrow -\boldsymbol{\gamma}$, in the third line we have exchanged the dummy variables of integration, and in the forth line have assumed that $\Psi^*_0(-\textbf k)=\Psi_0(\textbf k)$, which would hold for a rotationally-symmetric, aberration-free probe, and used the property of \cref{eq:b10} that
$F(\tau,\gamma,\hat{\boldsymbol{\tau}}\cdot\hat{\boldsymbol{\gamma}})=F(\gamma,\tau,\hat{\boldsymbol{\tau}}\cdot\hat{\boldsymbol{\gamma}})$. \Cref{eq:b11} shows that the inelastic diffraction pattern is inversion symmetric, and it then immediately follows that its centre of mass is zero:
\begin{align}
    \int d \textbf{k}_\perp \, k_x \sum_{m,m'} \mathcal{I}(\textbf{k}_\perp) &=
    \int d \textbf{k}_\perp \, (-k_x) \sum_{m,m'} \mathcal{I}(-\textbf{k}_\perp) = 
    - \int d \textbf{k}_\perp \, k_x \sum_{m,m'} \mathcal{I}(\textbf{k}_\perp) = 0
\end{align}
(the same result follows for the $x$-direction). Hence the first moment of an inelastic diffraction pattern would be zero if elastic scattering did not occur (assuming the phase object approximation and a rotationally-symmetric, aberration-free probe).


\begin{thebibliography}{10}
\expandafter\ifx\csname url\endcsname\relax
  \def\url#1{\texttt{#1}}\fi
\expandafter\ifx\csname urlprefix\endcsname\relax\def\urlprefix{URL }\fi
\expandafter\ifx\csname href\endcsname\relax
  \def\href#1#2{#2} \def\path#1{#1}\fi

\bibitem{jarausch2009four}
K.~Jarausch, P.~Thomas, D.~N. Leonard, R.~Twesten, C.~R. Booth,
  Four-dimensional {STEM-EELS}: Enabling nano-scale chemical tomography,
  Ultramicroscopy 109~(4) (2009) 326--337.
\newblock \href {https://doi.org/10.1016/j.ultramic.2008.12.012}
  {\path{doi:10.1016/j.ultramic.2008.12.012}}.

\bibitem{ophus_four-dimensional_2019}
C.~Ophus, Four-{Dimensional} {Scanning} {Transmission} {Electron} {Microscopy}
  ({4D}-{STEM}): {From} {Scanning} {Nanodiffraction} to {Ptychography} and
  {Beyond}, Microscopy and Microanalysis 25~(3) (2019) 563--582.
\newblock \href {https://doi.org/10.1017/S1431927619000497}
  {\path{doi:10.1017/S1431927619000497}}.

\bibitem{maclaren2020comparison}
I.~MacLaren, E.~Frutos-Myro, D.~McGrouther, S.~McFadzean, J.~K. Weiss,
  D.~Cosart, J.~Portillo, A.~Robins, S.~Nicolopoulos, E.~Nebot~del Busto,
  et~al., A comparison of a direct electron detector and a high-speed video
  camera for a scanning precession electron diffraction phase and orientation
  mapping, Microscopy and Microanalysis 26~(6) (2020) 1110--1116.
\newblock \href {https://doi.org/10.1017/S1431927620024411}
  {\path{doi:10.1017/S1431927620024411}}.

\bibitem{ophus2022automated}
C.~Ophus, S.~E. Zeltmann, A.~Bruefach, A.~Rakowski, B.~H. Savitzky, A.~M.
  Minor, M.~C. Scott, Automated crystal orientation mapping in {py4DSTEM} using
  sparse correlation matching, Microscopy and microanalysis 28~(2) (2022)
  390--403.
\newblock \href {https://doi.org/10.1017/S1431927622000101}
  {\path{doi:10.1017/S1431927622000101}}.

\bibitem{ozdol2015strain}
V.~Ozdol, C.~Gammer, X.~Jin, P.~Ercius, C.~Ophus, J.~Ciston, A.~Minor, Strain
  mapping at nanometer resolution using advanced nano-beam electron
  diffraction, Applied Physics Letters 106~(25) (2015).
\newblock \href {https://doi.org/10.1063/1.4922994}
  {\path{doi:10.1063/1.4922994}}.

\bibitem{shi2022uncovering}
C.~Shi, M.~C. Cao, S.~M. Rehn, S.-H. Bae, J.~Kim, M.~R. Jones, D.~A. Muller,
  Y.~Han, Uncovering material deformations via machine learning combined with
  four-dimensional scanning transmission electron microscopy, npj Computational
  Materials 8~(1) (2022) 114.
\newblock \href {https://doi.org/10.1038/s41524-022-00793-9}
  {\path{doi:10.1038/s41524-022-00793-9}}.

\bibitem{tate2016high}
M.~W. Tate, P.~Purohit, D.~Chamberlain, K.~X. Nguyen, R.~Hovden, C.~S. Chang,
  P.~Deb, E.~Turgut, J.~T. Heron, D.~G. Schlom, et~al., High dynamic range
  pixel array detector for scanning transmission electron microscopy,
  Microscopy and Microanalysis 22~(1) (2016) 237--249.
\newblock \href {https://doi.org/10.1017/S1431927615015664}
  {\path{doi:10.1017/S1431927615015664}}.

\bibitem{krajnak2016pixelated}
M.~Krajnak, D.~McGrouther, D.~Maneuski, V.~O'Shea, S.~McVitie, Pixelated
  detectors and improved efficiency for magnetic imaging in stem differential
  phase contrast, Ultramicroscopy 165 (2016) 42--50.
\newblock \href {https://doi.org/10.1016/j.ultramic.2016.03.006}
  {\path{doi:10.1016/j.ultramic.2016.03.006}}.

\bibitem{da2022influence}
B.~C. da~Silva, Z.~S. Momtaz, L.~Bruas, J.-L. Rouvi{\'e}re, H.~Okuno,
  D.~Cooper, M.~I. Den-Hertog, The influence of illumination conditions in the
  measurement of built-in electric field at p--n junctions by {4D-STEM},
  Applied Physics Letters 121~(12) (2022) 123503.
\newblock \href {https://doi.org/10.1063/5.0104861}
  {\path{doi:10.1063/5.0104861}}.

\bibitem{muller2014atomic}
K.~M{\"u}ller, F.~F. Krause, A.~B{\'e}ch{\'e}, M.~Schowalter, V.~Galioit,
  S.~L{\"o}ffler, J.~Verbeeck, J.~Zweck, P.~Schattschneider, A.~Rosenauer,
  Atomic electric fields revealed by a quantum mechanical approach to electron
  picodiffraction, Nature Communications 5~(1) (2014) 5653.
\newblock \href {https://doi.org/10.1038/ncomms6653}
  {\path{doi:10.1038/ncomms6653}}.

\bibitem{hachtel2018sub}
J.~A. Hachtel, J.~C. Idrobo, M.~Chi, Sub-{\aa}ngstrom electric field
  measurements on a universal detector in a scanning transmission electron
  microscope, Advanced Structural and Chemical Imaging 4 (2018) 1--10.
\newblock \href {https://doi.org/10.1186/s40679-018-0059-4}
  {\path{doi:10.1186/s40679-018-0059-4}}.

\bibitem{yang2016simultaneous}
H.~Yang, R.~Rutte, L.~Jones, M.~Simson, R.~Sagawa, H.~Ryll, M.~Huth,
  T.~Pennycook, M.~Green, H.~Soltau, et~al., Simultaneous atomic-resolution
  electron ptychography and {Z-contrast} imaging of light and heavy elements in
  complex nanostructures, Nature Communications 7~(1) (2016) 12532.
\newblock \href {https://doi.org/10.1038/ncomms12532}
  {\path{doi:10.1038/ncomms12532}}.

\bibitem{jiang2018electron}
Y.~Jiang, Z.~Chen, Y.~Han, P.~Deb, H.~Gao, S.~Xie, P.~Purohit, M.~W. Tate,
  J.~Park, S.~M. Gruner, et~al., Electron ptychography of {2D} materials to
  deep sub-{\aa}ngstr{\"o}m resolution, Nature 559~(7714) (2018) 343--349.
\newblock \href {https://doi.org/10.1038/s41586-018-0298-5}
  {\path{doi:10.1038/s41586-018-0298-5}}.

\bibitem{chen2021electron}
Z.~Chen, Y.~Jiang, Y.-T. Shao, M.~E. Holtz, M.~Odstr{\v{c}}il,
  M.~Guizar-Sicairos, I.~Hanke, S.~Ganschow, D.~G. Schlom, D.~A. Muller,
  Electron ptychography achieves atomic-resolution limits set by lattice
  vibrations, Science 372~(6544) (2021) 826--831.
\newblock \href {https://doi.org/10.1126/science.abg2533}
  {\path{doi:10.1126/science.abg2533}}.

\bibitem{bustillo20214d}
K.~C. Bustillo, S.~E. Zeltmann, M.~Chen, J.~Donohue, J.~Ciston, C.~Ophus, A.~M.
  Minor, {4D-STEM} of beam-sensitive materials, Accounts of Chemical Research
  54~(11) (2021) 2543--2551.
\newblock \href {https://doi.org/10.1021/acs.accounts.1c00073}
  {\path{doi:10.1021/acs.accounts.1c00073}}.

\bibitem{muto2017high}
S.~Muto, M.~Ohtsuka, High-precision quantitative atomic-site-analysis of
  functional dopants in crystalline materials by electron-channelling-enhanced
  microanalysis, Progress in Crystal Growth and Characterization of Materials
  63~(2) (2017) 40--61.
\newblock \href {https://doi.org/10.1016/j.pcrysgrow.2017.02.001}
  {\path{doi:10.1016/j.pcrysgrow.2017.02.001}}.

\bibitem{midgley1995energy}
P.~Midgley, M.~Saunders, R.~Vincent, J.~Steeds, Energy-filtered convergent-beam
  diffraction: examples and future prospects, Ultramicroscopy 59~(1-4) (1995)
  1--13.
\newblock \href {https://doi.org/10.1016/0304-3991(95)00014-R}
  {\path{doi:10.1016/0304-3991(95)00014-R}}.

\bibitem{hage2017momentum}
F.~Hage, T.~Hardcastle, A.~Scott, R.~Brydson, Q.~Ramasse, Momentum- and
  space-resolved high-resolution electron energy loss spectroscopy of
  individual single-wall carbon nanotubes, Physical Review B 95~(19) (2017)
  195411.
\newblock \href {https://doi.org/10.1103/PhysRevB.95.195411}
  {\path{doi:10.1103/PhysRevB.95.195411}}.

\bibitem{qi2021four}
R.~Qi, N.~Li, J.~Du, R.~Shi, Y.~Huang, X.~Yang, L.~Liu, Z.~Xu, Q.~Dai, D.~Yu,
  et~al., Four-dimensional vibrational spectroscopy for nanoscale mapping of
  phonon dispersion in {BN} nanotubes, Nature Communications 12~(1) (2021)
  1179.
\newblock \href {https://doi.org/10.1038/s41467-021-21452-5}
  {\path{doi:10.1038/s41467-021-21452-5}}.

\bibitem{haas_momentum-_2022}
B.~Haas, C.~T. Koch, Momentum- and {Energy}-{Resolved} {STEM} at {Atomic}
  {Resolution}, Microscopy and Microanalysis 28~(S1) (2022) 406--408.
\newblock \href {https://doi.org/10.1017/S1431927622002343}
  {\path{doi:10.1017/S1431927622002343}}.

\bibitem{dwyer_multislice_2005}
C.~Dwyer, Multislice theory of fast electron scattering incorporating atomic
  inner-shell ionization, Ultramicroscopy 104~(2) (2005) 141--151.
\newblock \href {https://doi.org/10.1016/j.ultramic.2005.03.005}
  {\path{doi:10.1016/j.ultramic.2005.03.005}}.

\bibitem{dwyer2008multiple}
C.~Dwyer, S.~D. Findlay, L.~J. Allen, Multiple elastic scattering of core-loss
  electrons in atomic resolution imaging, Physical Review B 77~(18) (2008)
  184107.
\newblock \href {https://doi.org/10.1103/PhysRevB.77.184107}
  {\path{doi:10.1103/PhysRevB.77.184107}}.

\bibitem{brown_linear-scaling_2019}
H.~G. Brown, J.~Ciston, C.~Ophus, Linear-scaling algorithm for rapid
  computation of inelastic transitions in the presence of multiple electron
  scattering, Physical Review Research 1~(3) (2019) 033186, publisher: American
  Physical Society.
\newblock \href {https://doi.org/10.1103/PhysRevResearch.1.033186}
  {\path{doi:10.1103/PhysRevResearch.1.033186}}.

\bibitem{muller2017measurement}
K.~M{\"u}ller-Caspary, F.~F. Krause, T.~Grieb, S.~L{\"o}ffler, M.~Schowalter,
  A.~B{\'e}ch{\'e}, V.~Galioit, D.~Marquardt, J.~Zweck, P.~Schattschneider,
  et~al., Measurement of atomic electric fields and charge densities from
  average momentum transfers using scanning transmission electron microscopy,
  Ultramicroscopy 178 (2017) 62--80.
\newblock \href {https://doi.org/10.1016/j.ultramic.2016.05.004}
  {\path{doi:10.1016/j.ultramic.2016.05.004}}.

\bibitem{close_towards_2015}
R.~Close, Z.~Chen, N.~Shibata, S.~D. Findlay, Towards quantitative,
  atomic-resolution reconstruction of the electrostatic potential via
  differential phase contrast using electrons, Ultramicroscopy 159 (2015)
  124--137.
\newblock \href {https://doi.org/10.1016/j.ultramic.2015.09.002}
  {\path{doi:10.1016/j.ultramic.2015.09.002}}.

\bibitem{seki2017quantitative}
T.~Seki, G.~S{\'a}nchez-Santolino, R.~Ishikawa, S.~D. Findlay, Y.~Ikuhara,
  N.~Shibata, Quantitative electric field mapping in thin specimens using a
  segmented detector: Revisiting the transfer function for differential phase
  contrast, Ultramicroscopy 182 (2017) 258--263.

\bibitem{dalfonso_three-dimensional_2008}
A.~J. D’Alfonso, E.~C. Cosgriff, S.~D. Findlay, G.~Behan, A.~I. Kirkland,
  P.~D. Nellist, L.~J. Allen, Three-dimensional imaging in double
  aberration-corrected scanning confocal electron microscopy, {Part} {II}:
  {Inelastic} scattering, Ultramicroscopy 108~(12) (2008) 1567--1578.
\newblock \href {https://doi.org/10.1016/j.ultramic.2008.05.007}
  {\path{doi:10.1016/j.ultramic.2008.05.007}}.

\bibitem{coene_inelastic_1990}
W.~Coene, D.~Van~Dyck, Inelastic scattering of high-energy electrons in real
  space, Ultramicroscopy 33~(4) (1990) 261--267.
\newblock \href {https://doi.org/10.1016/0304-3991(90)90043-L}
  {\path{doi:10.1016/0304-3991(90)90043-L}}.

\bibitem{wang2011spatial}
X.~Wang, P.~Li, M.~Malac, R.~Lockwood, A.~Meldrum, The spatial distribution of
  silicon {NC}s and erbium ion clusters by simultaneous high-resolution energy
  filtered and {Z}-contrast {STEM} and transmission electron tomography,
  Physica Status Solidi C 8~(3) (2011) 1038--1043.
\newblock \href {https://doi.org/10.1002/pssc.201000393}
  {\path{doi:10.1002/pssc.201000393}}.

\bibitem{xin2014there}
H.~L. Xin, C.~Dwyer, D.~A. Muller, Is there a {S}tobbs factor in
  atomic-resolution {STEM-EELS} mapping?, Ultramicroscopy 139 (2014) 38--46.
\newblock \href {https://doi.org/10.1016/j.ultramic.2014.01.006}
  {\path{doi:10.1016/j.ultramic.2014.01.006}}.

\bibitem{oxley2000atomic}
M.~Oxley, L.~Allen, Atomic scattering factors for {K}-shell and {L}-shell
  ionization by fast electrons, Acta Crystallographica Section A: Foundations
  of Crystallography 56~(5) (2000) 470--490.
\newblock \href {https://doi.org/10.1107/s0108767300007078}
  {\path{doi:10.1107/s0108767300007078}}.

\bibitem{pyms}
H.~Brown, py\_multislice, \url{https://github.com/HamishGBrown/py_multislice}.

\bibitem{lazic_phase_2016}
I.~Lazić, E.~G.~T. Bosch, S.~Lazar, Phase contrast {STEM} for thin samples:
  {Integrated} differential phase contrast, Ultramicroscopy 160 (2016)
  265--280.
\newblock \href {https://doi.org/10.1016/j.ultramic.2015.10.011}
  {\path{doi:10.1016/j.ultramic.2015.10.011}}.

\bibitem{howie1963inelastic}
A.~Howie, Inelastic scattering of electrons by crystals. {I}. the theory of
  small-angle in elastic scattering, Proceedings of the Royal Society of
  London. Series A. Mathematical and Physical Sciences 271~(1345) (1963)
  268--287.
\newblock \href {https://doi.org/10.1098/rspa.1963.0017}
  {\path{doi:10.1098/rspa.1963.0017}}.

\bibitem{brown_addressing_2016}
H.~Brown, A.~D'Alfonso, B.~Forbes, L.~Allen, Addressing preservation of elastic
  contrast in energy-filtered transmission electron microscopy, Ultramicroscopy
  160 (2016) 90--97.
\newblock \href {https://doi.org/10.1016/j.ultramic.2015.10.001}
  {\path{doi:10.1016/j.ultramic.2015.10.001}}.

\bibitem{Kottler:07}
C.~Kottler, C.~David, F.~Pfeiffer, O.~Bunk, A two-directional approach for
  grating based differential phase contrast imaging using hard x-rays, Opt.
  Express 15~(3) (2007) 1175--1181.
\newblock \href {https://doi.org/10.1364/OE.15.001175}
  {\path{doi:10.1364/OE.15.001175}}.

\bibitem{beyer2020influence}
A.~Beyer, F.~F. Krause, H.~L. Robert, S.~Firoozabadi, T.~Grieb,
  P.~K{\"u}kelhan, D.~Heimes, M.~Schowalter, K.~M{\"u}ller-Caspary,
  A.~Rosenauer, et~al., Influence of plasmon excitations on atomic-resolution
  quantitative {4D} scanning transmission electron microscopy, Scientific
  reports 10~(1) (2020) 17890.
\newblock \href {https://doi.org/10.1038/s41598-020-74434-w}
  {\path{doi:10.1038/s41598-020-74434-w}}.

\bibitem{winkler2020direct}
F.~Winkler, J.~Barthel, R.~E. Dunin-Borkowski, K.~M{\"u}ller-Caspary, Direct
  measurement of electrostatic potentials at the atomic scale: A conceptual
  comparison between electron holography and scanning transmission electron
  microscopy, Ultramicroscopy 210 (2020) 112926.
\newblock \href {https://doi.org/10.1016/j.ultramic.2019.112926}
  {\path{doi:10.1016/j.ultramic.2019.112926}}.

\bibitem{hillyard1993annular}
S.~Hillyard, R.~F. Loane, J.~Silcox, Annular dark-field imaging: resolution and
  thickness effects, Ultramicroscopy 49~(1-4) (1993) 14--25.
\newblock \href {https://doi.org/10.1016/0304-3991(93)90209-G}
  {\path{doi:10.1016/0304-3991(93)90209-G}}.

\bibitem{dwyer2003scattering}
C.~Dwyer, J.~Etheridge, Scattering of Å-scale electron probes in silicon,
  Ultramicroscopy 96~(3-4) (2003) 343--360.
\newblock \href {https://doi.org/10.1016/S0304-3991(03)00100-1}
  {\path{doi:10.1016/S0304-3991(03)00100-1}}.

\bibitem{allen2003lattice}
L.~Allen, S.~Findlay, M.~Oxley, C.~Rossouw, Lattice-resolution contrast from a
  focused coherent electron probe. {Part I}, Ultramicroscopy 96~(1) (2003)
  47--63.
\newblock \href {https://doi.org/10.1016/S0304-3991(02)00380-7}
  {\path{doi:10.1016/S0304-3991(02)00380-7}}.

\bibitem{voyles2004depth}
P.~Voyles, D.~Muller, E.~Kirkland, Depth-dependent imaging of individual dopant
  atoms in silicon, Microscopy and Microanalysis 10~(2) (2004) 291--300.
\newblock \href {https://doi.org/0.1017/S1431927604040012}
  {\path{doi:0.1017/S1431927604040012}}.

\bibitem{egerton2011electron}
R.~F. Egerton, Electron energy-loss spectroscopy in the electron microscope,
  Springer Science \& Business Media, 2011.

\bibitem{seki2018theoretical}
T.~Seki, Y.~Ikuhara, N.~Shibata, Theoretical framework of statistical noise in
  scanning transmission electron microscopy, Ultramicroscopy 193 (2018)
  118--125.

\end{thebibliography}








\section*{Supplementary material}

# Supplementary Material for "Differential phase contrast from electrons that cause inner shell ionization"

Michael Deimetry[a], Timothy C. Petersen[b], Hamish G. Brown[c], Matthew Weyland[b], Scott D. Findlay[a,*]

[a]*School of Physics and Astronomy, Monash University, Clayton, Victoria, 3800, Australia*
[b]*Monash Centre for Electron Microscopy, Monash University, Clayton, Victoria, 3800, Australia*
[c]*Ian Holmes Imaging Center, University of Melbourne, Parkville, Victoria, 3052, Australia*

## 1. Derivation of Eq. (11)

To evaluate Eq. (10), we use Parseval's theorem,

$$\int_{-\infty}^{\infty} d\mathbf{r}_\perp \, |H_{fi}(\mathbf{r}_\perp)|^2 = \int_{-\infty}^{\infty} d\mathbf{k}_\perp \, |H_{fi}(\mathbf{k}_\perp)|^2 \,, \tag{1}$$

and set $\boldsymbol{\tau} = \boldsymbol{\gamma} = \mathbf{k}_\perp$ in (A.3) to give

$$\begin{aligned}T_{\ell'}(n\ell, \varepsilon) &= \int d\mathbf{k}_\perp \sum_{m,m'} |H_{n\ell m \to \varepsilon \ell' m'}(\mathbf{k}_\perp)|^2 \\ &= \frac{q_e^4}{(4\pi^2 \varepsilon_0)^2} \frac{(2\ell'+1)(2\ell+1)}{(4\pi)^2} \sum_{\ell''=0}^{\infty} \begin{pmatrix} \ell' & \ell'' & \ell \\ 0 & 0 & 0 \end{pmatrix}^2 (2\ell''+1) \int \frac{d\mathbf{k}_\perp}{k^4} R_{l',l'',l}(k)^2 \end{aligned} \tag{2}$$

where we have used the fact that $P_{\ell''}(\hat{\mathbf{k}} \cdot \hat{\mathbf{k}}) = 1$. We focus on the weighted radial wave function overlap integral $R_{l',l'',l}(k)$ in Eq. (8), reproduced here for convenience:

$$R_{l',l'',l}(k) = \int_0^\infty dr \, P_{\varepsilon \ell'}(r) j_{\ell''}(kr) P_{nl}(r) \,. \tag{3}$$

To proceed with this integral we need to obtain expressions for the bound and continuum radial wave functions.

An analytic form for the bound wave function can be obtained using variational methods [42]. The form chosen is given by the Slater-type orbital

$$P_{n^*\ell}(r) = N r^{n^*} e^{-\zeta r} \,, \tag{4}$$

---

*Corresponding author

*Email address:* scott.findlay@monash.edu (Scott D. Findlay)

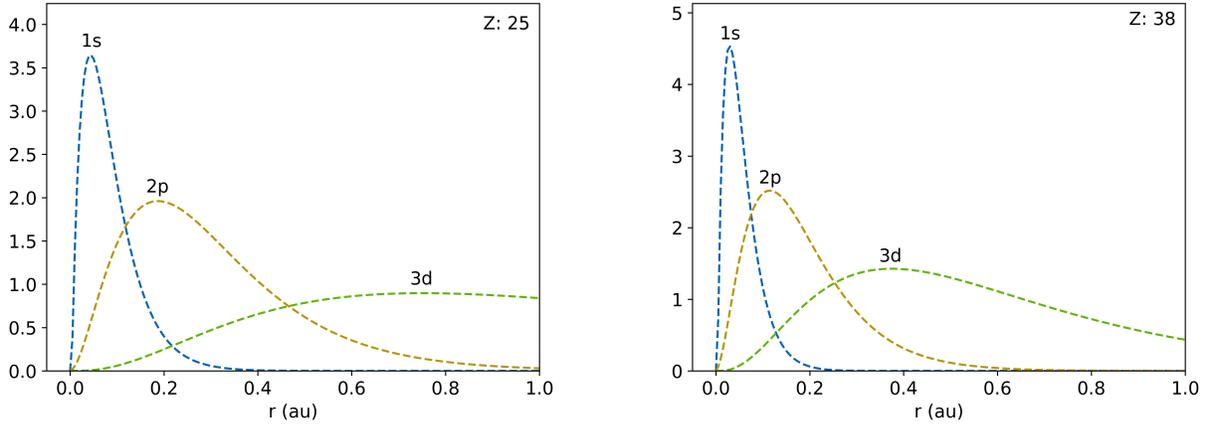

Figure 1: Bound radial wave functions for various quantum numbers.

where $N = (2\zeta)^{n^*}(2\zeta/\Gamma(2n^* + 1))^{1/2}$ is the normalization, $r$ is the distance from the nucleus in atomic units and $\zeta = (Z - s)/n^*$ where $Z$ is the atomic number, $n^*$ is a modified principal quantum number (which equals $n$ if $n \leq 3$) and $s$ is a so-called shielding factor. The remainder of this argument will only follow for bound states which are nodeless with non-negative range, for instance, $1s$, $2p$ or $3d$. These are illustrated in Fig. 1, where it is seen that they are only non-zero in close vicinity of the origin, consistent with the decaying exponential term in Eq. (4).

The unbound radial wave functions, $P_{\varepsilon\ell'}$, satisfy the differential equation [43]

$$\left[-\frac{d^2}{dr^2} + \frac{\ell'(\ell' + 1)}{r^2} + V_{\varepsilon\ell'}(r)\right] P_{\varepsilon\ell'}(r) = \varepsilon P_{\varepsilon\ell'}(r) \ . \tag{5}$$

Here $V_{\varepsilon\ell'}$ is the radial potential of the atom after an electron has been ejected with energy $\varepsilon$ above the ionization threshold given now in Rydberg units to reduce notational clutter. Since the radial norm of $P_{\varepsilon\ell'}(r)$ is not finite, we adopt the normalization convention [43]

$$\int_0^\infty P_{\varepsilon\ell'}(r) P_{\varepsilon'\ell'}(r) dr = \delta(\varepsilon - \varepsilon') \ . \tag{6}$$

Since the bound state wavefunctions are localised near the origin in $r$, as seen in Fig. 1, for the purpose of evaluating $R_{l',l'',l}(k)$ in Eq. (3) we are primarily interested in the form of $P_{\varepsilon\ell'}(r)$ for small $r$. In this regime, if $\ell'$ is sufficiently large the centrifugal barrier term $\frac{\ell'(\ell'+1)}{r^2}$ will dominate over the potential term $V_{\varepsilon\ell'}(r)$ and we can approximate Eq. (5) to

$$\left[-\frac{d^2}{dr^2} + \frac{\ell'(\ell' + 1)}{r^2}\right] P_{\varepsilon\ell'}(r) = \varepsilon P_{\varepsilon\ell'}(r) \ , \tag{7}$$

which has general solution

$$P_{\varepsilon\ell'}(r) \approx \sqrt{r} J_\nu(r\sqrt{\varepsilon}) c_1 + \sqrt{r} Y_\nu(r\sqrt{\varepsilon}) c_2 \ , \tag{8}$$

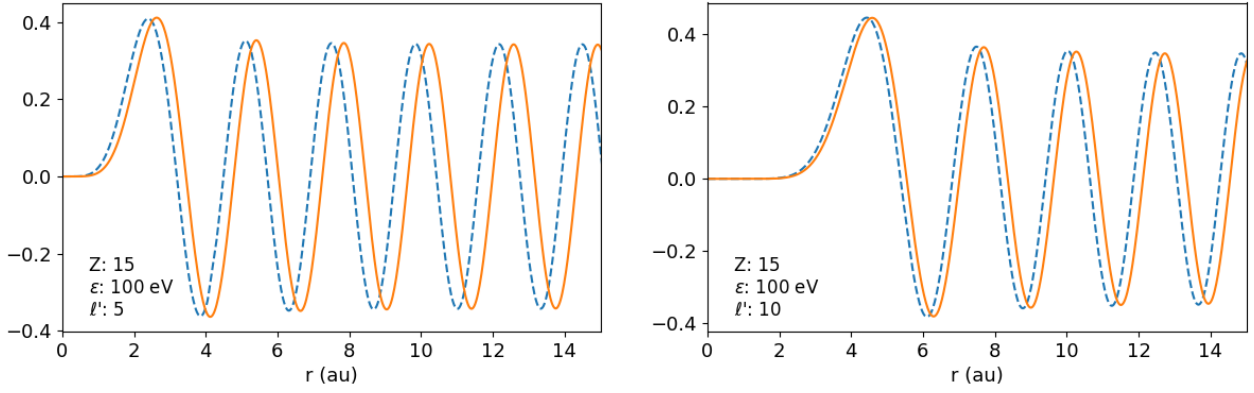

Figure 2: Continuum wave function obtained by solving Eq. (5) (dotted) in comparison with Eq. (10) (solid) which neglects the potential. For large $\ell'$ the centrifugal barrier dominates. This result is weakly dependent on $Z$ and $\varepsilon$.

where $J$ and $Y$ denote Bessel functions of the first and second kinds respectively, the subscript denotes the order and $c$ denote arbitrary constants. To satisfy the boundary condition

$$P_{\varepsilon\ell'}(r) \to 0 \quad \text{as} \quad r \to 0 \tag{9}$$

it follows that

$$P_{\varepsilon\ell'}(r) \approx c\sqrt{r}J_{\ell'+1/2}(r\sqrt{\varepsilon}) \ . \tag{10}$$

Using the large $x$ asymptotic result that [44]

$$J_\nu(x) \sim \sqrt{\frac{2}{\pi x}}\cos(x - \delta_\nu) \ , \tag{11}$$

it can be shown that the choice $c = 1/\sqrt{2}$ satisfies the normalisation condition of Eq. (6).

Some examples of the approximate continuum radial wave function $P_{\varepsilon\ell'}(r)$ in Eq. (10) in comparison with the true solution are illustrated in Fig. 2. As expected, increasing the centrifugal barrier by increasing $\ell'$ reduces the relative effect of the atom potential, bringing the approximate solution closer to the exact solution.

Being interested in the small $r$ regime, we simplify the radial wave functions integral further using the asymptotic approximation [44]

$$J_\nu(r) \sim \frac{e^{\nu - \left(\nu + \frac{1}{2}\right)\log(\nu) + \nu\log(r/2)}}{\sqrt{2\pi}} + \mathcal{O}(1/\nu) \ , \tag{12}$$

to yield

$$P_{\varepsilon\ell'}(r) \approx \sqrt{\frac{r}{2}}\frac{e^{\nu - \left(\nu + \frac{1}{2}\right)\log(\nu) + \nu\log(r\sqrt{\varepsilon}/2)}}{\sqrt{2\pi}} \ . \tag{13}$$

3

where $\nu = \ell' + 1/2$. Although we have analytic forms for all the factors in the integrand in Eq. (3), the resultant integral still does not admit an analytic soluion. We therefore make further approximations. We begin by applying the Cauchy-Schwarz inequality to obtain

$$|R(k)|^2 = \left|\int_0^\infty dr\, P_{\varepsilon\ell'}(r) j_{\ell''}(kr) P_{n^*\ell}(r)\right|^2 \leq \int_0^\infty dr\, |P_{\varepsilon\ell'}(r) P_{n^*\ell}(r)|^2 \int_0^\infty dr\, j_{\ell''}(kr)^2 \text{ for } k > 0,\tag{14}$$

which is a crude bound but, as we will see, is sufficient to imply an exponential fall-off.

Substituting Eq. (4) and Eq. (13) into the first integral on the right in Eq. (14) gives

$$\int_0^\infty dr\, |P_{\varepsilon\ell'}(r) P_{n^*\ell}(r)|^2 = \frac{N^2}{\pi} \frac{e^{2\ell'+1} \varepsilon^{\ell'+\frac{1}{2}} \Gamma(2\ell' + 2n^* + 3)}{(2\ell'+1)^{2(\ell'+1)} 2^{2(\ell'+n^*+2)} \zeta^{2\ell'+2n^*+3}} \tag{15}$$

The integral of the spherical Bessel function in Eq. (14) is straightforward and evaluates to

$$\int_0^\infty dr\, j_{\ell''}(kr)^2 = \frac{\pi}{2k(2\ell''+1)} \text{ if } k > 0. \tag{16}$$

Combining Eq. (14) with Eq. (2), we obtain

$$T_{\ell'}(n\ell, \varepsilon) \lesssim \frac{N^2 \pi}{12} \frac{q_e^4}{(4\pi^2\varepsilon_0)^2} \frac{(2\ell'+1)(2\ell+1)}{(4\pi)^2} \sum_{\ell''} \begin{pmatrix} \ell' & \ell'' & \ell \\ 0 & 0 & 0 \end{pmatrix}^2 \frac{e^{2\ell'+1} \varepsilon^{\ell'+\frac{1}{2}} \Gamma(2\ell' + 2n^* + 3)}{k_z^3 (2\ell'+1)^{2(\ell'+1)} 2^{2(\ell'+n^*+2)} \zeta^{2\ell'+2n^*+3}}. \tag{17}$$

Taking the ratio of this with the $\ell' = 0$ term and then taking the logarithm gives

$$\log_{10}\left[\frac{T_{\ell'}(n\ell,\varepsilon)}{T_0(n\ell,\varepsilon)}\right] \approx 2\ell' \log_{10}(e/2) + \log_{10}\left(\frac{\Gamma(2\ell'+2n^*+3)}{(1+2\ell')^{1+2\ell'}} \sum_{\ell''} \begin{pmatrix} \ell' & \ell'' & \ell \\ 0 & 0 & 0 \end{pmatrix}^2\right)$$
$$+ \ell' \log_{10} \varepsilon - 2\ell' \log_{10}(\zeta) + \text{other terms}. \tag{18}$$

We reiterate $\varepsilon$ in this equation is in Rydbergs while it is in electron volts in the main manuscript. This result elucidates the $\varepsilon$ and Z dependence observed in n Fig. 4. The first terms on the right hand side of Eq. (18) dictate the rate at which the fall-off occurs. We indeed see increasing $\varepsilon$ slows down the fall-off (by increasing or decreasing the gradient of the approximately linear behaviour in $\ell'$) while the converse is true for Z. This is made clearer when the relevant terms are grouped as follows:

$$\text{gradient} \approx 2\left[\log_{10} e + \frac{1}{2}\log_{10}\varepsilon - \log_{10}(\zeta)\right] \tag{19}$$

The second term on the right in Eq. (18) is inconsequential to this main result since it only creates minor deviations from the otherwise linear dependence on $\ell'$ in the logarithmic form of Eq. (18),

4

i.e. the sequence $T_{\ell'}(n\ell, \varepsilon)$ decreases exponentially with $\ell'$. The absence of $k_z$ is in line with expectations obtained from numerical simulations since the dependence is expected to be very weak. The approximations made here have produced an expression which has entirely neglected it.

We finally note that substituting $\varepsilon \to |\varepsilon - V_0|$ can better approximate the continuum wave function near the origin where the bound wave function admits a majority of its support. Through empirical considerations, a value of $V_0 = Z^{2/7}$ has been found to work well.



\end{document}